\documentclass[times,twocolumn]{aastex631}

\usepackage{comment}

\shorttitle{CCC in NGC~3627}
\shortauthors{Maeda et al.}
\graphicspath{{./}{figures/}}
\begin{document}

\title{Galactic structure dependence of cloud-cloud collisions driven star formation in the barred galaxy NGC~3627}

\author[0000-0002-8868-1255]{Fumiya Maeda}
\affiliation{Research Center for Physics and Mathematics, Osaka Electro-Communication University, 18-8 Hatsucho, Neyagawa, Osaka, 572-8530,  Japan}

\author[0000-0003-3844-1517]{Kouji Ohta}
\affiliation{Department of Astronomy, Kyoto University, Kitashirakawa-Oiwake-Cho, Sakyo-ku, Kyoto 606-8502, Japan}

\author[0000-0002-1639-1515]{Fumi Egusa}
\affiliation{Institute of Astronomy, Graduate School of Science, The University of Tokyo, 2-21-1 Osawa, Mitaka, Tokyo 181-0015, Japan}

\author[0000-0002-2107-1460]{Yusuke Fujimoto}
\affiliation{Department of Computer Science and Engineering, University of Aizu, Tsuruga Ikki-machi, Aizu-Wakamatsu, Fukushima, 965-8580, Japan}

\author[0000-0003-3990-1204]{Masato I.N. Kobayashi}
\affiliation{I. Physikalisches Institut, Universit\"{a}t zu K\"{o}ln, Z\"{u}lpicher Str 77, D-50937 K\"{o}ln, Germany
}

\author[0009-0004-6347-0613]{Shin Inoue}
\affiliation{Department of Astronomy, Kyoto University Sakyo-ku, Kyoto 606-8502, Japan}

\author{Asao Habe}
\affiliation{Graduate School of Science, Hokkaido University, Kita 10 Nishi 8, Kita-ku, Sapporo, Hokkaido 060-0810, Japan}

%\author[0000-0003-3983-5438]{Yoshihisa Asada}
%\affiliation{Department of Astronomy, Kyoto University Sakyo-ku, Kyoto 606-8502, Japan}



\begin{abstract}
While cloud-cloud collisions (CCCs) have been proposed as a mechanism for triggering massive star formation, it is suggested that higher collision velocities ($v_{\rm col}$) and lower GMC mass ($M_{\rm GMC}$) or/and density ($\Sigma_{\rm GMC}$) tend to suppress star formation. In this study, we choose the nearby barred galaxy NGC 3627 to examine the SFR and SFE of a colliding GMC ($m^\star_{\rm CCC}$ and $\epsilon_{\rm CCC}$) and explore the connections between $m^\star_{\rm CCC}$ and $\epsilon_{\rm CCC}$, $M_{\rm GMC}$($\Sigma_{\rm GMC}$) and $v_{\rm col}$, and galactic structures (disk, bar, and bar-end). 
Using ALMA CO(2--1) data (60~pc resolution), we estimated $v_{\rm col}$ within 500~pc apertures, based on line-of-sight GMC velocities, assuming random motion in a two-dimensional plane.
We extracted apertures where at least 0.1 collisions occur per 1 Myr, identifying them as regions dominated by CCC-driven star formation, and then calculated $m^\star_{\rm CCC}$ and $\epsilon_{\rm CCC}$ using attenuation-corrected H$\alpha$ data from VLT MUSE.
We found that both $m^\star_{\rm CCC}$ and $\epsilon_{\rm CCC}$ are lower in the bar (median values: $10^{3.84}~M_\odot$ and $0.18~\%$), and higher in the bar-end ($10^{4.89}~M_\odot$ and $1.10~\%$) compared to the disk ($10^{4.28}~M_\odot$ and $0.75~\%$). Furthermore, we found that structural differences within the parameter space of $v_{\rm col}$ and $M_{\rm GMC}$($\Sigma_{\rm GMC}$), with higher $M_{\rm GMC}$($\Sigma_{\rm GMC}$) in the bar-end and higher $v_{\rm col}$ in the bar compared to the disk, lead to higher star formation activity in the bar-end and lower activity in the bar. Our results support the scenario that variations in CCC properties across different galactic structures can explain the observed differences in SFE on a kpc scale within a disk galaxy.
\end{abstract}


\section{Introduction} \label{sec: intro}

To understand star formation in disk galaxies, it is crucial to uncover how large-scale galactic structures (e.g., center, bar, bar-end, and arm) affect the star formation of giant molecular clouds (GMCs). Recent kpc-scale observations reveal that star formation activity varies with galactic structure \citep[e.g.,][]{Handa_bar_1991,Momose_NGC4303_2010,Muraoka_NGC2903_2016,pan_variation_2017,law_submillimeter_2018,yajima_co_2019,Maeda_2023ApJ}. For instance, \citet{Maeda_2023ApJ} found that star formation efficiency (SFE) on a kpc scale—the ratio of star formation rate (SFR) to molecular gas surface density—is statistically lower in bars and higher in centers and bar-ends compared to spiral arms. They also observed a negative correlation between SFE and CO velocity width, with low SFE and high-velocity width in bars, and high SFE and low-velocity width in bar-ends. These findings suggest that dynamical processes driven by the galactic structures significantly influence the star formation processes of GMCs, leading to variations in kpc-scale SFEs.

Several mechanisms have been proposed to explain the lower  SFE in the bar region, particularly in relation to the large velocity dispersion. Bar-driven strong shocks and shear can inhibit the growth of gravitational instabilities and/or destroy GMCs, ultimately suppressing star formation \citep{tubbs_inhibition_1982,athanassoula_existence_1992,downes_co_1996,Reynaud_Downes_bar_1998,Zurita_2004AA,Kim_2012ApJ,Meidt_2013ApJ,emsellem_interplay_2015,renaud_environmental_2015,Kim_2024ApJ}. Observational studies further suggest that the vigorous motions in the bar region make it difficult for dense gas to survive, causing molecular gas to exist primarily as diffuse gas \citep[e.g.,][]{sorai_properties_2012, Muraoka_NGC2903_2016, maeda_a_large_2020}.

This paper focuses on an alternative scenario in which galactic structures may influence the properties of cloud-cloud collisions (CCCs) themselves. CCCs are proposed as a key process that compresses molecular gas at the shock front, triggering massive star formation \citep[e.g.,][]{Habe1992PASJ,Fukui2014ApJ...780}. Theoretical studies suggest that CCC-driven star formation may account for a few 10 to 50\% of the total star formation in a disk galaxy \citep[e.g.,][]{tan_star_2000,Kobayashi_2018PASJ,Horie_2024MNRAS}. 
Sub-parsec scale hydrodynamical simulations of CCCs have suggested that star formation strongly depends on the collision velocity and cloud mass (and/or density) \citep[e.g.,][]{takahira_cloud-cloud_2014, takahira_formation_2018, Sakre_2023MNRAS}. The higher the collision velocity and the lower the mass(density), the more star formation is suppressed. When the mass is fixed, faster CCCs can shorten the gas accretion phase, suppressing cloud core growth and massive star formation. At a given collision velocity, higher mass (i.e., larger size) results in a longer collision duration. Additionally, higher density increases the amount of surrounding gas available for accretion to the core. CCC observations in the Milky Way show similar dependencies \citep{Enokiya_2021PASJ, Fukui_2021PASJ}.

Recent observations and simulations suggest that variations in the collision velocity and mass(density) of colliding GMCs across different galactic structures may explain the observed differences in SFE on a kpc scale. Observations of GMCs in the barred galaxy NGC~1300 have reported that collision velocities may be higher in the bar region than in the arm region \citep{Maeda_CCC_2021}, which may contribute to star formation suppression in the bar. Although estimated collision velocities in the bar-end are similar to those in the bar region, star formation in the bar-end region is enhanced, likely due to the larger GMC mass(density) reported in the bar-end compared to the bar \citep{Maeda_GMC_2020MNRAS}. Supporting these findings, parsec-scale hydrodynamical simulations of barred galaxies indicate that GMC collision velocities in the bar region are higher than in the arm regions due to the violent gas motions driven by the bar potential \citep{fujimoto_giant_2014,fujimoto_environmental_2014,fujimoto_fast_2020} and high-density GMCs exist in the bar-end regions due to the gas inflows from both the bar and arm  \citep{emsellem_interplay_2015}.

Studies of the connection between CCC velocities and galactic structures in nearby disk galaxies have been limited to NGC 1300 \citep{Maeda_CCC_2021}. Focusing on the barred galaxy NGC~3627, this paper aims to quantitatively investigate the SFR and SFE of colliding GMCs, explore how GMC mass(surface density) and collision velocity are linked to galactic structures, and examine how these relationships impact the SFR and SFE of colliding GMCs.  NGC~3627 is a strongly barred galaxy at a distance of 11.32 Mpc \citep{Tully_2009AJ}, with an inclination angle of $i =57.3^\circ$ \citep{Lang_2020ApJ}, which is favorable for observing velocity signatures. The molecular gas surface density at the bar and bar-end regions on a kpc scale is the highest among the nearby barred galaxies \citep[$\sim 100~M_\odot\rm pc^{-2}$;][]{Maeda_2023ApJ}, suggesting a high number density of GMCs and frequent collisions. The shock tracer CH$_3$OH is detected in both the bar and bar-end regions \citep[][Watanabe et al. in prep]{Watanabe_2019ApJS}, and CO(2--1) observations at a 60 pc scale by  \citet{leroy_phangsalma_2021} reveal numerous double peak line profiles in these regions. The SFE on a kpc scale varies by about an order of magnitude between the bar and bar-end \citep{Maeda_2023ApJ}, providing an opportunity to observe how differences in CCC properties contribute to variations in star formation activity across galactic structures.

Direct observations of CCCs in NGC 3627, similar to those conducted in the Milky Way \citep[see review][]{Fukui_2021PASJ}, would be ideal. However, identifying CCCs, which requires an angular resolution of a few parsecs, is challenging to achieve across the whole galaxy even with the Atacama Large Millimeter/submillimeter Array (ALMA). Therefore, using a GMC catalog identified at a 60~pc scale, we estimate the collision velocity of GMCs from their line-of-sight velocities by assuming random motion in a two-dimensional plane. In Section \ref{sec: method}, we describe how to derive the collision velocity and SFR and SFE of a colliding GMC in detail. The data we used and the reduction process are presented in Section 3. In Section 4, we examine the SFR and SFE of colliding GMCs and explore the connections between the SFR and SFE of colliding GMCs, GMC mass(surface density) and collision velocity, and galactic structures. Section 5 investigates the uncertainties and provides interpretations of our results and comparisons with other studies. Finally, Section 6 presents a summary of this study.

\section{Methodology} \label{sec: method}
%Based on the CCC star formation model \citep[e.g.][]{tan_star_2000}, SFR surface density in an aperture (a subregion) is expressed as
%\begin{equation}
%    \Sigma_{\rm SFR}^{\rm ap} = \epsilon f_{\rm sf} f_{\rm CCC} n_{\rm GMC} \bar{M}_{\rm GMC},
%\end{equation}
%according to Tan (2000).
In an aperture (a subregion) where the star formation process is dominated by CCCs, SFR surface density can be expressed as
\begin{equation}
    \Sigma_{\rm SFR}^{\rm ap} = \epsilon f_{\rm sf} \nu_{\rm CCC} n_{\rm GMC} \bar{M}_{\rm GMC},
\end{equation}
based on the CCC star formation model proposed by \citet{tan_star_2000}. Here $\epsilon$ is the total mass fraction of GMC gas converted to stars during a star-forming collision, $f_{\rm sf}$ is the fraction of cloud collisions that successfully lead to star formation, $\nu_{\rm CCC}$ is the collision frequency, $n_{\rm GMC}$ is the surface number density of GMCs in the aperture, and $\bar{M}_{\rm GMC}$ is the ensemble mean GMC mass in the aperture. A galaxy simulation by \citet{fujimoto_environmental_2014} successfully reproduced the low SFE observed in the bar region by keeping $\epsilon$ constant and reducing  $f_{\rm sf}$ for high-speed collisions. However, the $\epsilon$ and $f_{\rm sf}$ are not independent, and  $\epsilon$ is not constant but varies with the collision velocity and the cloud mass/density as shown in sub-parsec simulations \citep{takahira_formation_2018}. Since it is difficult to estimate $\epsilon$ and $f_{\rm sf}$  separately from observations, in this study we define $\epsilon f_{\rm sf}$  as SFE per CCC, $\epsilon_{\rm CCC}$, and investigate the collision velocity and GMC mass/density dependence of $\epsilon_{\rm CCC}$.

In this study, we focus on regions where the CCC-driven star formation is considered to dominate and estimate $\nu_{\rm CCC}$ from the GMC line-of-sight velocity by assuming that the GMCs are in random motion on a two-dimensional plane within a hexagonal aperture as shown in Figure 1. 
Due to the random-like motion of clouds induced by the elongated gas stream in the bar potential, the $\nu_{\rm CCC}$ determined by tracking GMCs generally agrees with those estimated under the assumption of random GMC motion in the bar and bar-end regions \citep{fujimoto_fast_2020, Maeda_CCC_2021}.

The collision velocity in the aperture is derived as 
\begin{equation}
     v_{\rm col} = \frac{\sqrt{2}}{\sin i} \sqrt{\frac{1}{N_{\rm GMC}} \sum_j^{N_{\rm GMC}} (v_{\rm los}^j - \bar{v}_{\rm los}  )^2 },
\end{equation}
where $v_{\rm los}^j$ is the line-of-sight velocity of the $j$-th GMC and $\bar{v}_{\rm los}$ is the mean line-of-sight velocity within the aperture, $N_{\rm GMC}$ is the number of GMCs in the aperture, and $i$ is the inclination of the disk galaxy. Then, the  $\nu_{\rm CCC}$ is derived as 
\begin{equation}
\label{nuccc}
\nu_{\rm CCC} = 2 \bar{R}_{\rm GMC}  n_{\rm GMC}  v_{\rm col}, 
\end{equation}
where $\bar{R}_{\rm GMC}$ is the ensemble mean radius of GMCs in aperture.
%, and   $n_{\rm GMC}$ is the GMC number surface density in the aperture. 

Using this $\nu_{\rm CCC}$, we can calculate the following physical quantities within an aperture: the number of collisions that occur per unit of time ($N_{\rm CCC}$), the total mass of stars formed per CCC ($m^\star_{\rm CCC}$), and the SFE per CCC ($\epsilon_{\rm CCC}$), given by 
\begin{eqnarray}
    \label{Nccc}
    N_{\rm CCC} &=& \nu_{\rm CCC} N_{\rm GMC},\\
    \label{mstar}
    m^\star_{\rm CCC} &=&\frac{{\Sigma}_{\rm SFR}^{\rm ap}}{\nu_{\rm CCC} n_{\rm GMC}},\\
    \label{epsilon}
    \epsilon_{\rm CCC}  &=& \frac{{\Sigma}_{\rm SFR}^{\rm ap}}{\nu_{\rm CCC} n_{\rm GMC} \bar{M}_{\rm GMC}} = \frac{m^\star_{\rm CCC}}{\bar{M}_{\rm GMC}} \label{eq: epsilion}.
\end{eqnarray}
%where $\bar{M}_{\rm GMC}$ and $\Sigma_{\rm SFR}^{\rm ap}$ are the mean GMC mass and mean SFR surface density in the aperture, respectively. 
Since $n_{\rm GMC} \bar{M}_{\rm GMC}$ represents the molecular gas surface density in the aperture, $\Sigma_{\rm mol}^{\rm ap}$, and $1/\nu_{\rm CCC}$ corresponds to the collision timescale, $t_{\rm CCC}$, Equation (\ref{eq: epsilion}) is equivalent to the ratio of $t_{\rm CCC}$ to the molecular gas depletion time in the aperture ($t_{\rm dep} = \Sigma_{\rm mol}^{\rm ap}/\Sigma_{\rm SFR}^{\rm ap}$), i.e., $\epsilon_{\rm CCC} = t_{\rm CCC}/t_{\rm dep}$.

\begin{figure}[t!]
 \begin{center}
  \includegraphics[width=75mm]{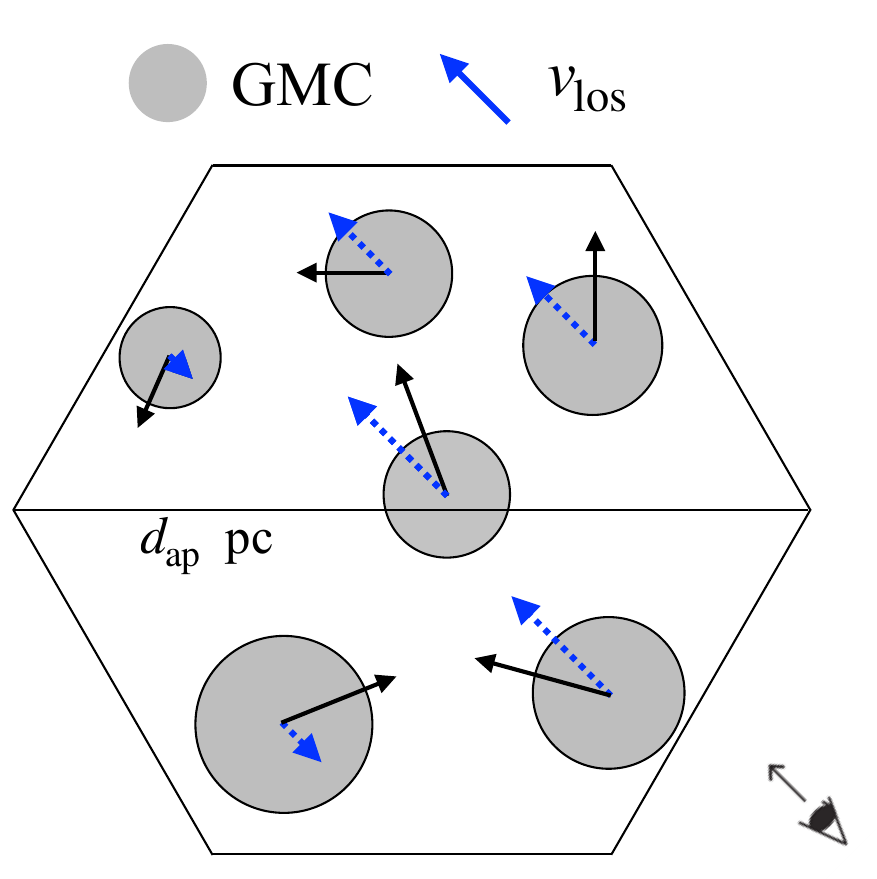} 
\caption{Schematic illustration showing the distribution of GMCs within the hexagonal aperture. The gray-filled circles represent the GMCs. The black vector and blue dotted vector are the velocity vector of a GMC and its line-of-sight velocity component. }\label{fig:sample}
 \end{center}
\end{figure}

\begin{figure*}[t!]
 \begin{center}
  \includegraphics[width=125mm]{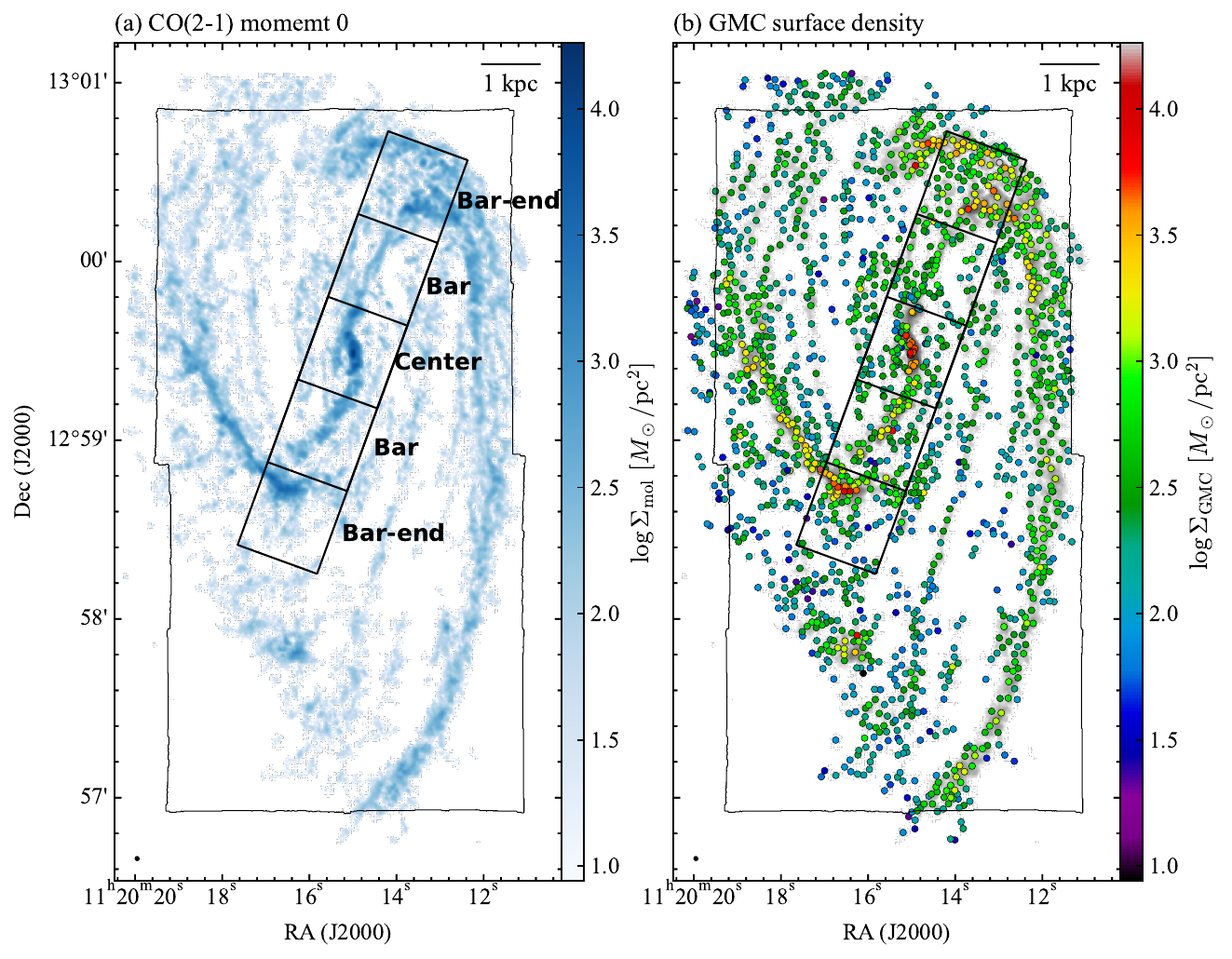} 
 \end{center}
\caption{
(a) CO(2--1) moment 0 map of NGC~3627. 
The color shows the molecular gas surface density ($\Sigma_{\rm mol}$).
We adopt a constant Milky Way $\alpha_{\rm CO}$ of $4.35~M_\odot~\rm (K~km~s^{-1}~pc^{-2})^{-1}$ and $R_{\rm 21} = 0.65$. 
The black-filled circle at the lower left corner represents the beam size.
The black rectangles represent the boundaries of the center, the bar, and the bar-end regions defined by \citet{Maeda_2023ApJ}.
The black frame is the FOV of the MUSE observations.
(b) GMC distribution in NGC~3627. The color represents the molecular gas surface density of the GMC. The background gray image is the panel (a).  }\label{fig:GMC_pos}
\end{figure*}

\begin{figure*}[t]
 \begin{center}
  \includegraphics[width=170mm]{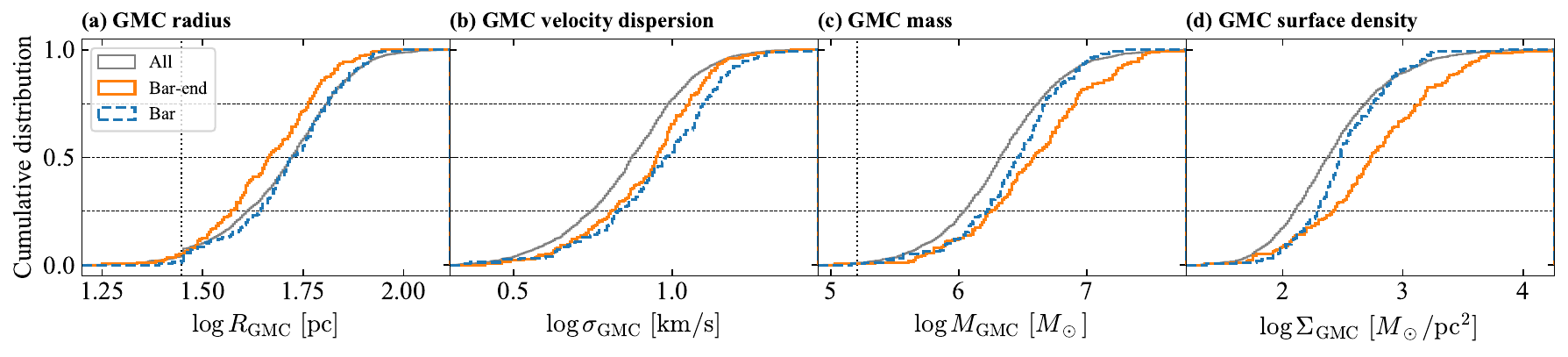} 
 \end{center}
\caption{Properties of the GMCs in NGC~3627. (a) The normalized cumulative distribution function of GMC radius in the whole (gray solid line), bar-end (orange bold line), and bar region (blue dotted line), respectively. The regions of bar-end and bar are represented in Figure~\ref{fig:GMC_pos}.
(b)-(d) Same as in panel (a), but for GMC velocity dispersion, luminosity mass, and surface density, respectively. The vertical lines in panel (a) and (c) show the resolution limit of the radius (28~pc) and sensitivity limit of mass ($1.6 \times 10^5~M_\odot$). The resolution limit of the velocity dispersion is $0.99~\rm km~s^{-1}$.}
\label{fig:GMC_prop_hist}
\end{figure*}

\section{Data and Reduction}

\subsection{Catalog of Giant Molecular Clouds}
\subsubsection{Data reduction for CO(2--1)}
We constructed a CO(2--1) data cube at a 60 pc scale using archival data observed with the ALMA under project 2015.1.00956.S. The observations were part of the   Physics at High-Angular resolution in Nearby GalaxieS
 (PHANGS)-ALMA project \citep{leroy_phangsalma_2021}, which achieved mapping of the disks of 90 nearby massive star-forming galaxies in CO(2--1) with an angular resolution of approximately $1^{\prime\prime}$. The field of view (FOV) of the NGC~3627 observations is about $2.^\prime5 \times 4.^\prime5$, covering most of the galaxy's inner disk as shown in Figure~\ref{fig:GMC_pos}(a). The observations were conducted using the 12-m, 7-m, and Total Power (TP) arrays. For both the 12 m and 7 m arrays data, we utilized calibrated visibility data provided by the East Asian ALMA Regional Center. The calibration of the TP data was performed using the Common Astronomy Software Application (CASA) package \citep{CASA_2022PASP..134k4501C} version 4.7.0, following the script provided by the observatory.

We used CASA version 6.4.0 to reconstruct the data cube. After concatenating the visibility data sets of 12 m and 7 m arrays data, we used the \verb|tclean| task with the \verb|multi-scale| deconvolver \citep{Kepley_2020PASP} for recovering extended emissions. Here, we considered scales of $0^{\prime\prime}$, $1^{\prime\prime}$, $2.^{\prime\prime}5$, $5^{\prime\prime}$, and $10^{\prime\prime}$. In \verb|tclean| task, we applied Briggs weighting with a robust of 0.5 and used the \verb|auto-multithresh| procedure to automatically identify regions containing emission in the dirty and residual images. We continued the deconvolution process until the intensity of the residual image attained the $\sim 3\sigma$ noise level. The calibrated TP data was imaged in \verb|sdimaging| task.
Then, using the CASA task \verb|feather|, the interferometric cube was combined with the TP cube.
The resultant angular resolution is $1.^{\prime\prime}09 \times 0.^{\prime\prime}98$, corresponding to $\rm 60~pc \times 54~pc$, with a position angle of $38^\circ.6$. The median rms noise ($\sigma_{\rm rms}$) of the data cube is $\rm 6.11~mJy~beam^{-1}$, corresponding to 132~mK in 2.5~$\rm km~s^{-1}$ bin. Finally, we applied the primary beam correction on the output-restored image cube.

\subsubsection{GMC identification}\label{sec: GMC identification}

We identified the GMCs in NGC~3627 using the PYCPROPS Python package \citep{Rosolowsky_GMC_2021MNRAS}. This package implements the CPROPS algorithm, originally described by \citet{Rosolowsky_Leroy_2006}, and leverages the fast dendrogram algorithm provided by the ASTRODENDRO package for emission segmentation. Although the GMCs in NGC 3627 were previously identified using PYCPROPS with a CO(2--1) data cube at 90 pc resolution \citep{Rosolowsky_GMC_2021MNRAS}, in this study, we re-identified them at a higher resolution of 60 pc.

Below, we briefly outline the identification process and the derivation of the GMC properties. We began by identifying regions with significant emissions within the data cube. Voxels were selected where the signal exceeded 4.0$\sigma_{\rm rms}$ in at least two adjacent velocity channels. These voxels were then expanded to include all adjacent pixels above 2.0$\sigma_{\rm rms}$. Using the PYCPROPS, we decomposed the regions with significant emissions into individual GMCs. The PYCPROPS parameters for GMC identification were set to match those used in \citet{Rosolowsky_GMC_2021MNRAS}. The process started with identifying local maxima, where the intensity ($T_{\rm max}$) was required to be at least 2.0$\sigma_{\rm rms}$ higher than the merging level ($T_{\rm merge}$). $T_{\rm merge}$ is defined as the highest value contour containing a given local maximum and one other neighboring peak. Additionally, the number of pixels above $T_{\rm merge}$ had to exceed a quarter of the beam size. Finally, the PYCPROPS assigned voxels around each local maximum to the corresponding peak using the watershed algorithm, with each local maximum being designated as an individual independent GMC.

\subsubsection{Properties of GMCs}

Here we summarize the properties of GMCs.
The total number of GMCs is 1803, excluding  those with $T_{\rm max} < 4.0\sigma_{\rm rms}$. The distribution of the GMCs is shown in Figure~\ref{fig:GMC_pos}(b). The line-of-sight velocity of a GMC, $v_{\rm los}$, is defined as the intensity-weighted mean velocity.  PYCPROPS corrects for the sensitivity by extrapolating GMC properties to those we would expect to measure with perfect sensitivity (i.e. 0 K) and the resolution by deconvolution for the beam and channel width (see \citealt{Rosolowsky_Leroy_2006} for details). The radius is defined as $R_{\rm GMC} = \eta \sqrt{\sigma_{\rm maj}\sigma_{\rm min}}$, where $\sigma_{\rm maj}$ and $\sigma_{\rm min}$ are the extrapolated and deconvolved intensity-weighted second moments along the major and minor axis, respectively. The coefficient $\eta$ is set to be $\sqrt{2\ln2}$ by assuming the surface brightness of the clouds follows a two-dimensional Gaussian \citep[see][]{Rosolowsky_GMC_2021MNRAS}. Thus, the $R_{\rm GMC}$ of a GMC with a size comparable to the beam size is equal to half of the geometric mean of the beam dimensions, which is 28~pc.
The cumulative distribution function (CDF) of the $R_{\rm GMC}$ is shown in Figure~\ref{fig:GMC_prop_hist}(a). The median $R_{\rm GMC}$ in whole, bar-end, and bar are 53, 46, and 53~pc, respectively. Here, the definitions of the bar-end and bar regions follow
\citet{Maeda_2023ApJ}, as  shown in Figure~\ref{fig:GMC_pos} and described in Section~\ref{sec: region mask}.

The velocity dispersion of a GMC is defined as $\sigma_{\rm GMC} = \sqrt{\sigma_{\rm ex}^2 - \sigma_{\rm chan}^2}$, where $\sigma_{\rm ex}$ is the extrapolated velocity dispersion. The $\sigma_{\rm chan}$ is the equivalent Gaussian width of a channel and is defined as $\sigma_{\rm chan} = W/\sqrt{2\pi} = 0.99~\rm km~s^{-1}$, where $W$ is the channel width of $2.5~\rm km~s^{-1}$. The median $\sigma_{\rm GMC}$ in whole, bar-end, and bar are 7.5, 9.0, and 9.5~$\rm km~s^{-1}$, respectively (Figure~\ref{fig:GMC_prop_hist}(b)).

The luminosity mass of a GMC is derived as $M_{\rm GMC} = \alpha_{\rm CO}L_{\rm CO(2-1)}/R_{21}$, where $L_{\rm CO(2-1)}$ is the extrapolated luminosity, $\alpha_{\rm CO}$ is a CO(1--0)-to-H$_2$ conversion factor, and $R_{21}$ is a CO(2--1)-to-CO(1--0) brightness temperature ratio.
In this study, we adopt a constant Milky Way $\alpha_{\rm CO}$ of $4.35~M_\odot~\rm (K~km~s^{-1}~pc^{-2})^{-1}$, including a factor of 1.36, to account for the presence of helium \citep{bolatto_conversion_2013}. We adopt $R_{\rm 21} = 0.65$ based on \citet{leroy_molecular_2013} and \citet{den_brok_new_2021}, measured at kpc scales. We discuss the effect of uncertainties in the $\alpha_{\rm CO}$ and $R_{\rm 21}$ on our results in Sections \ref{sec: CO(2--1)/CO(1--0) line ratio} and \ref{sec: conversion factor}.
The mass detection limit is estimated to be $1.6 \times 10^5~M_\odot$, based on the assumption that the minimum detectable GMC has a peak intensity equivalent to $4\sigma_{\rm rms}$, a size comparable to the beam size, and a velocity width equal to twice the channel width. The median $M_{\rm GMC}$ in whole, bar-end, and bar are $10^{6.33}$, $10^{6.59}$, and $10^{6.47}$~$M_\odot$, respectively (Figure~\ref{fig:GMC_prop_hist}(c)).

The molecular gas surface density of a GMC is defined as $\Sigma_{\rm GMC} = M_{\rm GMC}/(\pi R_{\rm GMC}^2$). 
The median $\Sigma_{\rm GMC}$ across the whole region is 241~$M_\odot~\rm pc^{-2}$.
As illustrated in Figures~\ref{fig:GMC_pos}(a) and \ref{fig:GMC_prop_hist}(d), the $\Sigma_{\rm GMC}$ in the bar-end regions is systematically higher than that in the bar region. The median $\Sigma_{\rm GMC}$ in the bar-end and bar are 557 and 308~$M_\odot~\rm pc^{-2}$, respectively. This difference arises because the bar-end regions tend to have systematically larger $M_{\rm GMC}$ and smaller $R_{\rm GMC}$ compared to the bar region (Figures~\ref{fig:GMC_prop_hist}(a) and (d)).

\begin{deluxetable*}{ccccccccccc}
\tablecaption{Physical properties within the apertures ($d_{\rm ap} = 500~\rm pc$) with $N_{\rm CCC} > 0.1~\rm Myr^{-1}$.\label{tab: physical prop 500pc}}
\tablewidth{0pt}
\tablehead{ 
   Region & \# &$\log \bar{M}_{\rm GMC}$ & $\log \bar{\Sigma}_{\rm GMC}$ & $\log \Sigma_{\rm SFR}^{\rm ap}$ & $n_{\rm GMC}$ & $v_{\rm col}$ & $\nu_{\rm CCC}$ & $N_{\rm CCC}$ & $\log  m^\star_{\rm CCC} $  & $\epsilon_{\rm CCC}$\\ 
      & & ($M_\odot$) & ($M_\odot~\rm pc^{-2}$) & ($M_\odot~\rm yr^{-1}~kpc^{-2}$)& ($\rm kpc^{-2}$) & ($\rm km~s^{-1}$) & ($\rm Gyr^{-1}$) & ($\rm Myr^{-1}$) & ($M_\odot$) & (\%) 
 }
\decimalcolnumbers
\startdata
All & 406 & $6.49_{-0.22}^{+0.18}$ & $2.59_{-0.24}^{+0.25}$ & $-1.77_{-0.37}^{+0.42}$ & $20.0_{-3.3}^{+5.8}$ & $19.4_{-4.5}^{+10.7}$ & $38.1_{-10.1}^{+21.0}$ & $0.22_{-0.07}^{+0.21}$ & $4.30_{-0.44}^{+0.45}$ & $0.73_{-0.44}^{+0.88}$ \\
Disk & 301 & $6.42_{-0.21}^{+0.18}$ & $2.52_{-0.24}^{+0.25}$ & $-1.83_{-0.36}^{+0.38}$ & $20.0_{-3.3}^{+3.3}$ & $18.4_{-4.5}^{+7.6}$ & $36.5_{-9.6}^{+15.6}$ & $0.20_{-0.05}^{+0.13}$ & $4.28_{-0.42}^{+0.37}$ & $0.75_{-0.43}^{+0.86}$ \\
Bar-end & 51 & $6.81_{-0.27}^{+0.23}$ & $3.05_{-0.32}^{+0.23}$ & $-1.04_{-0.32}^{+0.28}$ & $23.3_{-3.3}^{+3.3}$ & $21.0_{-4.5}^{+9.1}$ & $46.1_{-12.1}^{+16.4}$ & $0.31_{-0.09}^{+0.13}$ & $4.89_{-0.32}^{+0.23}$ & $1.10_{-0.50}^{+1.97}$ \\
Bar & 54 & $6.54_{-0.10}^{+0.21}$ & $2.65_{-0.14}^{+0.11}$ & $-1.89_{-0.26}^{+0.38}$ & $23.3_{-6.7}^{+3.3}$ & $44.4_{-24.7}^{+18.3}$ & $101.1_{-64.3}^{+62.3}$ & $0.59_{-0.38}^{+0.64}$ & $3.84_{-0.33}^{+0.38}$ & $0.18_{-0.11}^{+0.39}$ \\
\enddata
\tablecomments{
Each physical property is noted as $M^{+D75}_{-D25}$, where $M$, $D25$, and $D75$ are the median, the distance to the 25th percentile from the median, and the distance to the 75th percentile from the median of the number distribution, respectively.}
\end{deluxetable*}

\begin{figure*}[t!]
 \begin{center}
  \includegraphics[width=125mm]{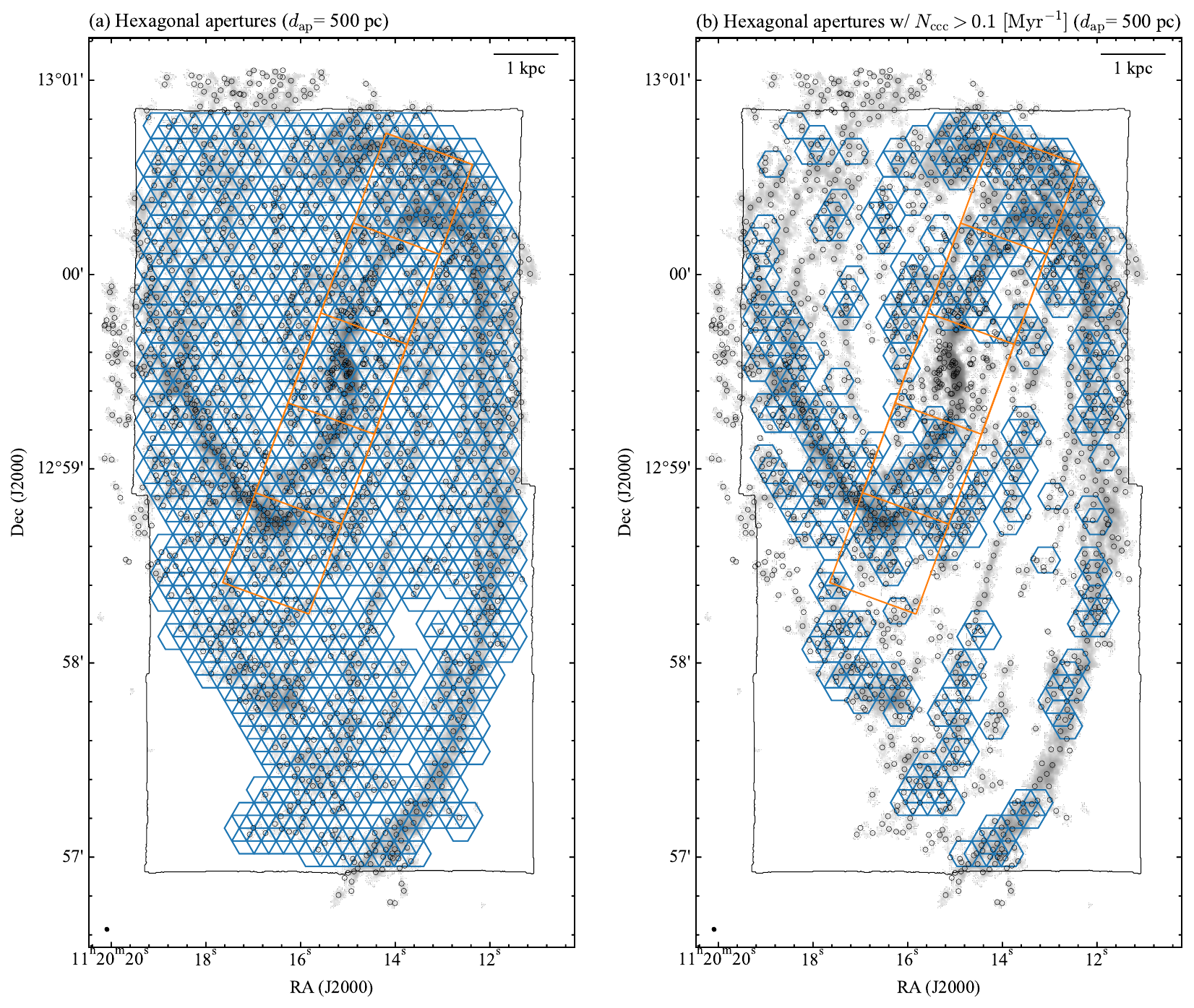} 
 \end{center}
\caption{(a) Position of hexagonal apertures (blue) with a size of $d_{\rm ap} = 500~\rm pc$ to cover all the GMCs (open circle) in the FoV of the H$\alpha$ image (black frame). The orange rectangles represent the boundaries of the center, the bar, and the
bar-end regions. (b) Apertures with $N_{\rm CCC} \geq 0.1~\rm Myr^{-1}$. Here, $N_{\rm CCC}$ is the number of collisions that occur per unit of time within an aperture.
 }\label{fig:hex_tile}

 \begin{center}
  \includegraphics[width=160mm]{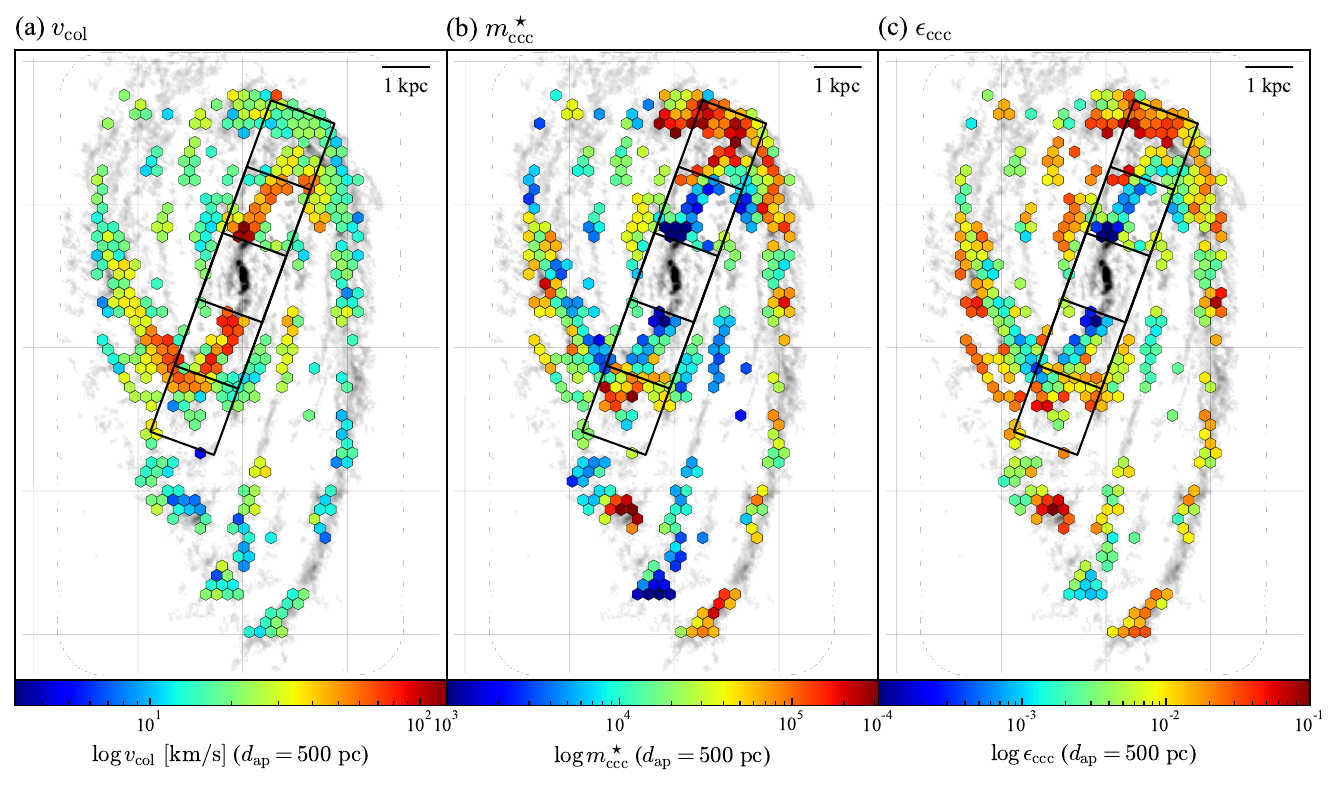} 
 \end{center}
\caption{Distribution of (a) the collision velocity ($v_{\rm col}$), (b) the total mass of stars formed per CCC ($m^\star_{\rm CCC}$), and (c) the SFE per CCC ($\epsilon_{\rm CCC}$) within an aperture in NGC~3627.
We used the apertures with $N_{\rm CCC} \geq 0.1~\rm Myr^{-1}$ (Figure~\ref{fig:hex_tile}(b)).  The black rectangles represent the boundaries of the center, the bar, and the bar-end regions. 
 }\label{fig:vcol_mstar_eccc_map}
\end{figure*}

\subsection{H$\alpha$}
\label{sec: Halpha}
As an SFR tracer, we used the H$\alpha$ image provided by the PHANGS-MUSE project \citep{Emsellem_phangs_muse2022}, which has the same resolution as the CO(2--1) data. The PHANGS-MUSE project provides fully calibrated data cubes and maps of 19 nearby galaxies including NGC~3627. From their data archive, we retrieved H$\alpha$ and H$\beta$ maps at $1.^{\prime\prime}05$ resolution. Note that [NII] lines around H$\alpha$ are fitted separately and that these maps are already corrected for the Milky Way foreground extinction. The observed flux ratio of H$\alpha$ to H$\beta$ is used to correct for internal extinction to the H$\alpha$ emission. We adopted parameters in Table 2 of \citet{Calzetti_2001PASP} for this calculation, and excluded pixels where signal-to-noise ratio (S/N) $< 3$  for either of the two emissions. If the calculated extinction becomes negative, no correction is applied. The median values of correction are 0.72 mag.

The $\Sigma_{\rm SFR}^{\rm ap}$ is derived as 
\begin{equation}
    \left( \frac{\Sigma_{\rm SFR}^{\rm ap}}{M_\odot~\rm yr^{-1}~kpc^{-2}} \right)
     = 5.37 \times 10^{-42} \left( \frac{L_{\rm H\alpha}^{\rm ap}}{\rm erg~s^{-1}} \right) \left( \frac{A_{\rm ap}}{\rm kpc^{2}} \right)^{-1} ,
\end{equation}
where  $L_{\rm H\alpha}^{\rm ap}$ is the attenuation corrected H$\alpha$ luminosity in the aperture, and $A_{\rm ap}$ is the deprojected area of the aperture. The coefficient of $5.37 \times 10^{-42}~M_\odot~\rm yr^{-1}~erg^{-1}~s$ is the conversion factor from H$\alpha$ luminosity to the SFR obtained by \citet{Murphy_2011ApJ}. They adopted Kroupa IMF \citep{Kroupa_IMF_2001}. 

Because NGC~3627 is classified as a low-ionization nuclear emission-line region (LINER)/type 2 Seyfert galaxy \citep[e.g.,][]{Ho_1997ApJS,Filho_2000ApJS}, most of the H$\alpha$ emission in the center contains contamination from the AGN. Using the BPT diagram, i.e., $\rm [OIII]/H\beta$ versus $\rm [N II]/H\alpha$ diagram, most of the H$\alpha$ there were flagged as AGN \citep{Emsellem_phangs_muse2022, Groves_2023MNRAS}. Therefore, the center region defined in Figure~\ref{fig:GMC_pos}(a) is not used in this study because of the large uncertainties of the SFR.

\subsection{Region mask}
\label{sec: region mask}
We adopted the definition of the center, bar, and the bar-end regions by \citet{Maeda_2023ApJ}. Figure~\ref{fig:GMC_pos} shows the definition as black rectangles.  \citet{Maeda_2023ApJ} defined a rectangle to enclose the elliptical stellar bar defined based on Spitzer 3.6~$\mu$m by \citet{herrera-endoqui_catalogue_2015} and then divided the rectangle into five equal boxes. The position angle of the bar is $160^\circ$. The length and width of the black rectangle are 8.0 and  1.5~kpc, respectively. Among the five defined boxes, the central box is referred to as the “center,” the two ends as “bar-ends,” and the boxes between them as the “bar”. The length of the rectangle was set to 1.25 times the major axis of the elliptical stellar bar to ensure that the SFR peak at the edge of the ellipse and surrounding region are included as the bar-end region.  The size of the center region is approximately four times the effective radius of the bulge \citep{Salo_2015ApJS..219....4S}, encompassing majority of the bulge light. Notably, most of the emissions flagged as AGN in the BPT diagram \citep{Emsellem_phangs_muse2022,Groves_2023MNRAS} are located within this center region.

\begin{figure*}[]
 \begin{center}
  \includegraphics[width=180mm]{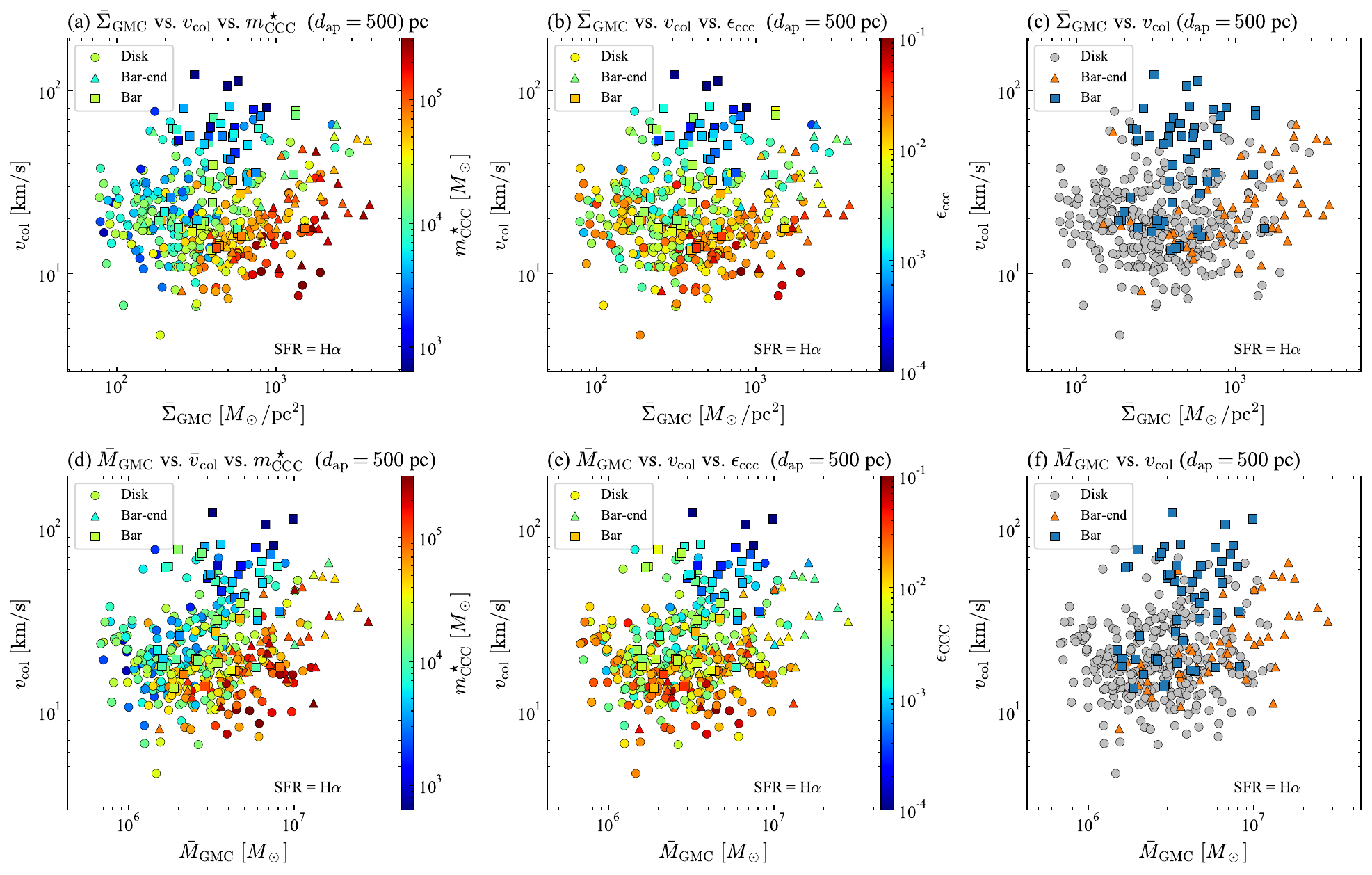} 
 \end{center}
\caption{Dependence of star formation on collision velocity and GMC surface density/mass in NGC~3627. SFR is derived from H$\alpha$ emission corrected for attenuation (Section~\ref{sec: Halpha}). 
(a) Dependence of the total mass of stars formed per CCC ($m^\star_{\rm CCC}$) on the $\bar{\Sigma}_{\rm GMC}$ and the $v_{\rm col}$.
(b) Dependence of the SFE per CCC ($\epsilon_{\rm CCC}$) on the $\bar{\Sigma}_{\rm GMC}$ and the $v_{\rm col}$. 
(c) Distribution of disk (grey circle), bar-end (orange triangle), and bar (blue square) apertures in the parameter space of $v_{\rm col}$ and $\bar{\Sigma}_{\rm GMC}$.
(d) - (f) Same as the panel (a)-(c), but using $\bar{M}_{\rm GMC}$ instead of  $\bar{\Sigma}_{\rm GMC}$.
 }\label{fig:main_result}
\end{figure*}

\section{Results} \label{sec: results}

\subsection{Aperture setting}
In this study, hexagonal apertures with a size (diameter) of $d_{\rm ap} = 500~\rm pc$ were positioned to cover all the GMCs in the FoV of the H$\alpha$ image as shown in Figure~\ref{fig:hex_tile}(a).  The centers of these apertures were spaced at intervals of $d_{\rm ap}/2 = 250~\rm pc$, representing a form of Nyquist sampling that minimizes arbitrariness in the aperture setting. The total number of apertures is 1157. The size of the aperture is roughly comparable to the width of the dust lane in the bar and arm region. 

After excluding 157 apertures that contained only one GMC, we calculated the $v_{\rm col}$, $\nu_{\rm CCC}$, $N_{\rm CCC}$, $m^\star_{\rm CCC}$ and $\epsilon_{\rm CCC}$.
To extract regions where the CCC-driven star formation is considered to be dominated, we then identified apertures with $N_{\rm CCC} \geq 0.1~\rm Myr^{-1}$ as regions where CCCs are considered to occur frequently.  The apertures in the center region were excluded from the analysis due to the inability to estimate $\Sigma_{\rm SFR}$ accurately because of the LINER activity (see Section~\ref{sec: Halpha}). As a result, 406 apertures were extracted from a total of 1157 apertures. Figure~\ref{fig:hex_tile}(b) illustrates the apertures used in our analysis, showing that apertures on the bar, bar-end, and spiral arms were selected. The number of apertures in the bar-end and bar region are 51 and 54, respectively. This figure shows that regions with strong CO(2--1) emissions generally correspond to regions where CCCs are frequent in NGC~3627.
The dependencies of our results on the $d_{\rm ap}$ and the aperture extraction criterion are discussed in Section \ref{sec: Uncertainties}.

\subsection{Physical properties within the apertures}

Table~\ref{tab: physical prop 500pc} shows the median values of the physical properties within the apertures ($d_{\rm ap} = 500~\rm pc$) with $N_{\rm CCC} > 0.1~\rm Myr^{-1}$ in each region. The scatter is defined as the distance from the 25th percentile to the 75th percentile (the so-called “interquartile range”, IQR). In this table, “Disk” refers to the subgroup of 301 apertures other than the bar-end and bar regions. Most of the apertures classified as “disk” are located within the spiral arms, although some are in the inter-arm regions as well.
Similar to Figures~\ref{fig:GMC_prop_hist}(c)-(d), 
the $\bar{M}_{\rm GMC}$ and ensemble mean GMC surface density in the aperture, $\bar{\Sigma}_{\rm GMC}$, are the highest in the bar-end region. The  $\Sigma_{\rm SFR}^{\rm ap}$ derived from H$\alpha$ emission in the bar-end region is about 6 times higher than that in the disk region.

Figure~\ref{fig:vcol_mstar_eccc_map}(a) shows the distribution of the collision velocity, $v_{\rm col}$. The median $v_{\rm col}$ in NGC~3627 is 19.4~$\rm km~s^{-1}$.
We find that the $v_{\rm col}$ in the bar region is higher, with a median of 44.4~$\rm km~s^{-1}$, compared to a median of 21.0~$\rm km~s^{-1}$ in the bar-end region. 
The calculated $\nu_{\rm CCC}$ and $N_{\rm CCC}$ are also higher in the bar region compared to the bar-end region. This is primarily due to the difference in the $v_{\rm col}$ because there is no significant difference in the $n_{\rm GMC}$ between the bar and bar-end regions. The collision timescale for a GMC, $t_{\rm CCC}$, is the inverse of the $\nu_{\rm CCC}$, which is 27.3, 9.89, and 21.7~Myr in the disk, bar, and bar-end regions, respectively.

Figure~\ref{fig:vcol_mstar_eccc_map}(b) and (c) show the distribution of the total mass of stars formed per CCC, $m^\star_{\rm CCC}$, and the SFE per CCC in the aperture, $\epsilon_{\rm CCC}$, respectively.  In NGC 3627, the median $m^\star_{\rm CCC}$ and $\epsilon_{\rm CCC}$ are $10^{4.30}~M_\odot$ and $0.73~\%$, respectively. Moreover, the median $m^\star_{\rm CCC}$ and $\epsilon_{\rm CCC}$ are lower in the bar region, at $10^{3.84}~M_\odot$ and $0.18~\%$, and higher in the bar-end region, at $10^{4.89}~M_\odot$ and $1.10~\%$, compared to $10^{4.28}~M_\odot$ and $0.75~\%$ in the disk region. These findings suggest that star formation activity driven by CCCs is suppressed in the bar region and enhanced in the bar-end region relative to the disk region in NGC~3627. We observe that both $m^\star_{\rm CCC}$ and $\epsilon_{\rm CCC}$ decrease along the bar from the bar-end toward the center, with $m^\star_{\rm CCC}$  declining from $\sim 10^{5.5}~M_\odot$ to $\sim 10^{3.5}~M_\odot$ and $\epsilon_{\rm CCC}$ decreasing from $\sim$10\% to $\sim$0.01\%.

\begin{figure*}[]
 \begin{center}
  \includegraphics[width=150mm]{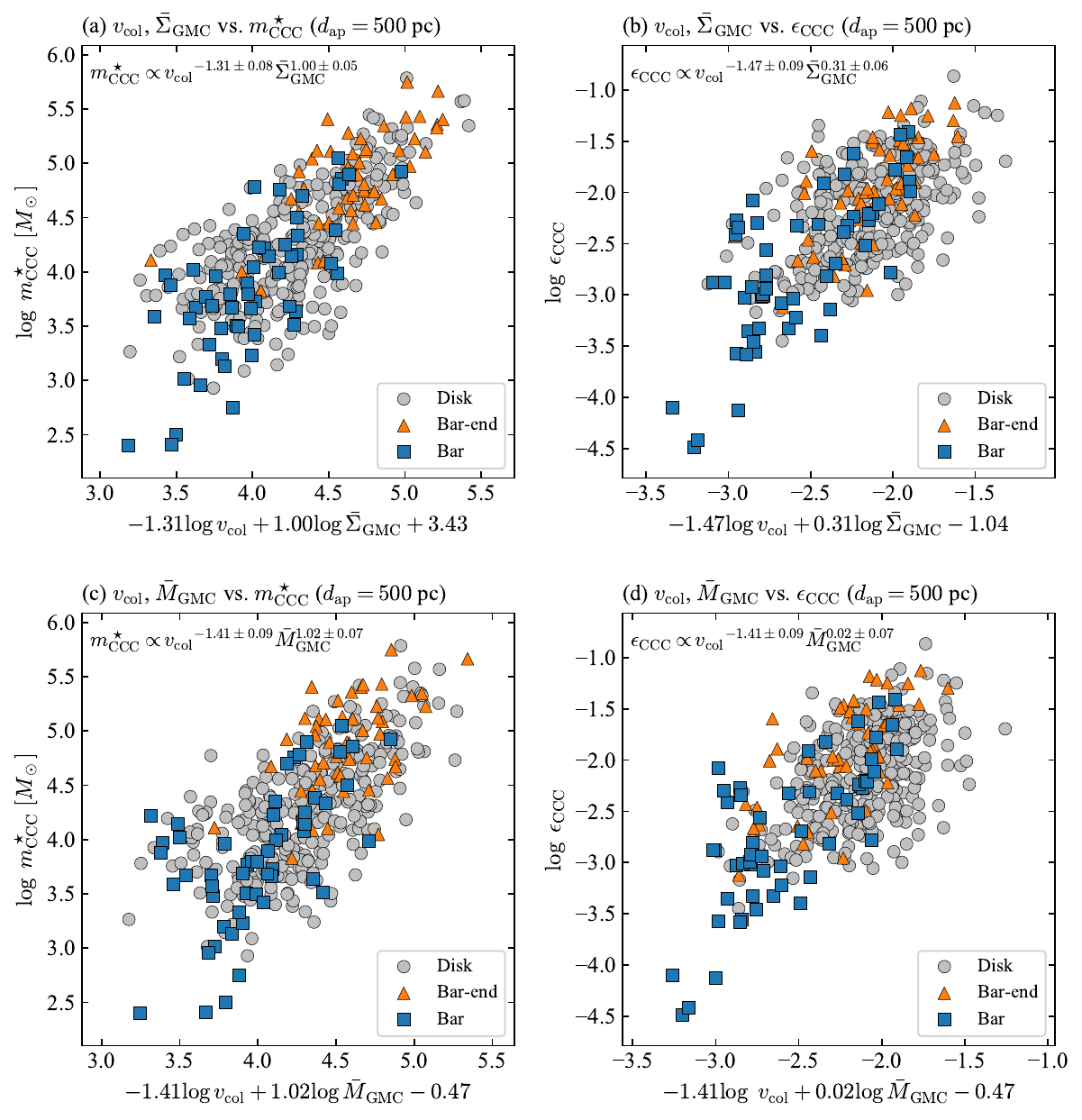} 
 \end{center}
\caption{ Fitting results of the dependence of $m^\star_{\rm CCC}$ and $\epsilon_{\rm CCC}$ on $v_{\rm col}$ and $\bar{\Sigma}_{\rm GMC}$ ($\bar{M}_{\rm GMC}$).
 }\label{fig:fitting}
\end{figure*}

\subsection{Connection between star formation, collision velocity and GMC surface density/mass, and galactic structures} \label{sec: dependence}
This section explores the connections between the SFR and SFE of the colliding GMC, collision velocity and GMC mass (or surface density), and galactic structures. 
Figure \ref{fig:main_result}(a) shows the dependence of $m^\star_{\rm CCC}$ on $v_{\rm col}$ and $\bar{\Sigma}_{\rm GMC}$.
The $m^\star_{\rm CCC}$ decreases with both increasing $v_{\rm col}$ and decreasing $\bar{\Sigma}_{\rm GMC}$.
When $v_{\rm col}$ exceeds $50~\rm km~s^{-1}$ and $\bar{\Sigma}_{\rm GMC}$ is below $300~M_\odot~\rm pc^{-2}$, $m^\star_{\rm CCC}$ generally remains under $10^4~M_\odot$.
Figure \ref{fig:main_result}(b) shows dependence of  $\epsilon_{\rm CCC}$ on $v_{\rm col}$ and $\bar{\Sigma}_{\rm GMC}$.
Similar to $m^\star_{\rm CCC}$, the $v_{\rm col}$ dependence of $\epsilon_{\rm CCC}$ is clearly seen, while the  $\bar{\Sigma}_{\rm GMC}$ dependence is weaker than that seen for $m^\star_{\rm CCC}$.
Note that, due to the calculation method, $m^\star_{\rm CCC}$ and $\epsilon_{\rm CCC}$ inherently tend to decrease as $v_{\rm col}$ increases. Specifically, $m^\star_{\rm CCC}$ and $\epsilon_{\rm CCC}$ are calculated using Equations (\ref{nuccc}),(\ref{mstar}) and (\ref{epsilon}), which inherently results in a tendency to be inversely proportional to $v_{\rm col}$.

Figure \ref{fig:main_result}(c) shows the distribution of disk, bar-end, and bar apertures in the parameter space of $v_{\rm col}$ and $\bar{\Sigma}_{\rm GMC}$. When comparing the bar and bar-end regions, apertures in the bar tend to exhibit lower $\bar{\Sigma}_{\rm GMC}$ and higher $v_{\rm col}$ (see also Table~\ref{tab: physical prop 500pc}). This distribution difference leads to lower $m^\star_{\rm CCC}$ and $\epsilon_{\rm CCC}$ in the bar, contrasting to higher $m^\star_{\rm CCC}$ and $\epsilon_{\rm CCC}$ in the bar-end region.

Figures \ref{fig:main_result}(d) and (e) are the same as panels (a) and (b) but for the dependence on $v_{\rm col}$ and $\bar{M}_{\rm GMC}$.
Similar to panel (a), $m^\star_{\rm CCC}$ decreases with decreasing $\bar{M}_{\rm GMC}$. However, the dependence of $\epsilon_{\rm CCC}$  on  $\bar{M}_{\rm GMC}$ at a fixed $v_{\rm col}$ is not observed.

To clarify the structural differences within the parameter space of $v_{\rm col}$ and $\bar{\Sigma}_{\rm GMC}$($\bar{M}_{\rm GMC}$),
we performed a fitting assuming that $m^\star_{\rm CCC}$ and $\epsilon_{\rm CCC}$ are proportional to a product of powers of $v_{\rm col}$ and $\bar{\Sigma}_{\rm GMC}$ ($\bar{M}_{\rm GMC}$). We used the \verb|curve_fit| function in Python’s SciPy package, which applies non-linear least squares to fit a function to the data.  Figure~\ref{fig:fitting} shows the fitting results, with coefficient values provided in Table~\ref{tab: fitting uncer}. As shown in Figure~\ref{fig:main_result}(a) and (b), both $m^\star_{\rm CCC}$ and  $\epsilon_{\rm CCC}$ decrease with both increasing $v_{\rm col}$ and decreasing $\bar{\Sigma}_{\rm GMC}$:
\begin{eqnarray}
    m^\star_{\rm CCC} \propto v_{\rm col}^{-1.31\pm0.08}\bar{\Sigma}_{\rm GMC}^{1.00\pm0.06}, \\
    \epsilon_{\rm CCC} \propto v_{\rm col}^{-1.47\pm0.09}\bar{\Sigma}_{\rm GMC}^{0.31\pm0.06}.     
\end{eqnarray}
As shown in Figure~\ref{fig:main_result}(d) and (e), $m^\star_{\rm CCC}$ depends on mass, whereas $\epsilon_{\rm CCC}$ is mass independent:
\begin{eqnarray}
    m^\star_{\rm CCC} \propto v_{\rm col}^{-1.41\pm0.09}\bar{M}_{\rm GMC}^{1.02\pm0.07}, \\
    \epsilon_{\rm CCC} \propto v_{\rm col}^{-1.41\pm0.09}\bar{M}_{\rm GMC}^{0.02\pm0.07}. 
\end{eqnarray}
Figure~\ref{fig:fitting} clearly show that the different structures have different distributions in the parameter space, with higher $\bar{\Sigma}_{\rm GMC}$ ($\bar{M}_{\rm GMC}$) in the bar-end regions and higher $v_{\rm col}$ in the bar regions compared to the disk region, leading to differences in star-forming activity across these structures.

Once again, note that the dependence of $m^\star_{\rm CCC}$ and $\epsilon_{\rm CCC}$ inherently results in a tendency to be inversely proportional to $v_{\rm col}$ by definition. However, the fact that the best-fit power of $v_{\rm col}$ is less than $-1$ indicates that $m^\star_{\rm CCC}$ and $\epsilon_{\rm CCC}$ exhibit a stronger dependence on $v_{\rm col}$ than what is expected from the definition.
This result may suggest that $v_{\rm col}$ is physically important for suppressing star formation.
Although $\epsilon_{\rm CCC}$ inherently tends to be inversely proportional to $\bar{M}_{\rm GMC}$, since it is calculated as $m^\star_{\rm CCC}/\bar{M}_{\rm GMC}$ (Equation~(\ref{eq: epsilion})), dependence on $\bar{M}_{\rm GMC}$ does not exhibit. This is because $m^\star_{\rm CCC}$ is proportional to $\bar{M}_{\rm GMC}$, which effectively cancels out the dependence of $\bar{M}_{\rm GMC}$ on $\epsilon_{\rm CCC}$.

\section{Discussions}
\subsection{Uncertainties} \label{sec: Uncertainties}
This subsection discusses the uncertainties in our results as presented in Section~\ref{sec: results}. The results of our study under different conditions are summarized in Tables~\ref{tab: uncertainties} and \ref{tab: fitting uncer}. While detailed analyses are provided below, our findings summarize that aperture size (Section~\ref{sec: Aperture size}) and the choice of SFR tracer (Section~\ref{sec: SFR}) contribute most to the uncertainties in our results. Nevertheless, our conclusions remain robust that structural differences within the parameter space of $v_{\rm col}$ and $M_{\rm GMC}$ ($\Sigma_{\rm GMC}$), with higher $M_{\rm GMC}$ ($\Sigma_{\rm GMC}$) in the bar-end and higher $v_{\rm col}$ in the bar compared to the disk, lead to higher star formation activity in the bar-end and lower activity in the bar.

\subsubsection{Aperture size}\label{sec: Aperture size}
We varied the $d_{\rm ap}$ from 300 to 700~pc to investigate the effect of the aperture size on our results, keeping all other conditions the same. As shown in Table~\ref{tab: uncertainties}, $\bar{M}_{\rm GMC}$, $\bar{\Sigma}_{\rm GMC}$, $\Sigma_{\rm SFR}^{\rm ap}$, and $v_{\rm col}$ are independent of $d_{\rm ap}$. However, the $n_{\rm GMC}$ decreases as $d_{\rm ap}$ increases, leading to a corresponding decrease in $\nu_{\rm CCC}$, $m^\star_{\rm CCC}$, and $\epsilon_{\rm CCC}$, resulting in the uncertainties of about a factor of 2. This occurs because the distribution of GMCs is not uniform in the galaxy. Additionally, because the dust lane (i.e., molecular gas) width is about 500 pc (Figure~\ref{fig:GMC_pos}),  a larger aperture will encompass regions where CO is not detected. Although the values of $m^\star_{\rm CCC}$ and $\epsilon_{\rm CCC}$ depend on the  $d_{\rm ap}$, fitting of the dependence of $m^\star_{\rm CCC}$ and $\epsilon_{\rm CCC}$ on the $v_{\rm col}$ and $\bar{\Sigma}_{\rm GMC}$ ($\bar{M}_{\rm GMC}$) are almost the same (Table~\ref{tab: fitting uncer}).

\subsubsection{Aperture extraction criterion}\label{sec: Aperture extraction criteria}
When extracting apertures, it is necessary to select regions where star formation driven by CCC is expected to be dominant. In Section~\ref{sec: results}, we imposed the condition of $N_{\rm CCC} = N_{\rm GMC}/t_{\rm CCC} \geq 0.1~\rm Myr^{-1}$ to include regions where, even if the $t_{\rm CCC}$ is long, a high GMC number density results in frequent CCCs within the aperture. However, this criterion includes regions where $t_{\rm CCC}$ is longer than the typical lifetime of a GMC of 30 Myr \citep[e.g.,][]{Kawamura_2009ApJS,Chevance_2020MNRAS,Kim_2022MNRAS.516.3006K,Demachi_2024PASJ}.
Here, to more strictly extract regions where CCC-driven star formation is considered to be dominant, we select regions with $t_{\rm CCC}$ smaller than  30 Myr. In this case, the number of apertures extracted is reduced by 40, 24, and 19\% in the disk, bar, and bar-end regions, respectively, due to the exclusion of apertures with low $v_{\rm col}$. This leads to a higher mean $N_{\rm CCC}$ (Table~\ref{tab: uncertainties}).  Consequently, the mean $m^\star_{\rm CCC}$ and $\epsilon_{\rm CCC}$ decrease by 30--40\%. The fitting of the dependence of $m^\star_{\rm CCC}$ and $\epsilon_{\rm CCC}$ on the $v_{\rm col}$ and $\bar{\Sigma}_{\rm GMC}$ ($\bar{M}_{\rm GMC}$) are almost the same (Table~\ref{tab: fitting uncer}).

In the case of $N_{\rm CCC} \geq  0.1~\rm Myr^{-1}$, the extracted apertures account for $\sim80~\%$ of the total SFR (excluding the center region). When applying the more stringent criterion of $t_{\rm CCC} < 30~\rm Myr$, this fraction decreases to $\sim60~\%$. These values exceed the $\sim10–50\%$ predicted by theoretical simulations \citep[e.g.,][]{Kobayashi_2018PASJ, Horie_2024MNRAS}. The higher fraction in NGC 3627 is reasonable since NGC~3627 has a high gas surface density and a strong bar structure,  which are not considered in these predictions. Notably, the bar-end region, where CCCs are expected to be frequent, accounts for as much as $\sim35~\%$ of the total SFR and likely contributes to the higher fraction.

\subsubsection{SFR}\label{sec: SFR}
To check the uncertainty of the choice of star formation tracer, we use the SFR obtained from GALEX far-ultraviolet (FUV) and Spitzer 24$\mu$m instead of attenuation-corrected H$\alpha$. The $\Sigma_{\rm SFR}$ can be calculated from a linear combination of  FUV and 24-$\mu$m intensities by
\citet{leroy_star_2008} as 
\begin{eqnarray}
    \label{eq: sfr ir}
    \left( \frac{\Sigma_{\rm SFR}}{M_\odot~\rm yr^{-1}~kpc^{-2}} \right) = [ 8.1 \times 10^{-2} \left( \frac{I_{\rm FUV}}{\rm MJy~sr^{-1}} \right) \nonumber \\ + 3.2 \times 10^{-3}   \left( \frac{I_{\rm 24 \mu m}}{\rm MJy~sr^{-1}} \right) ] \cos i,
\end{eqnarray}
where $I_{\rm FUV}$ and $I_{\rm 24 \mu m}$ are the FUV and 24~$\mu$m intensities, respectively. The first and second terms of this equation are unobscured and embedded SFR terms, respectively. These equations assume the Kroupa IMF \citep{Kroupa_IMF_2001}. 

We used GALEX FUV and Spitzer 24-$\mu$m archival data provided by Local Volume Legacy Survey \citep{Dale_LVL_2009ApJ}, which are available at \dataset[10.26131/IRSA414]{\doi{10.26131/IRSA414}}.
We used the background-subtracted FUV image with 5$^{\prime\prime}$ resolution and 24-$\mu$m image with 6$^{\prime\prime}$ resolution. Thus, after convolving the FUV image to 6$^{\prime\prime}$ resolution, the $\Sigma_{\rm SFR}$ image was created following Equation~(\ref{eq: sfr ir}). The $\Sigma_{\rm SFR}^{\rm ap}$ is derived from the mean value of the pixels in the aperture in the image.

The $\Sigma_{\rm SFR}^{\rm ap}$ values derived from H$\alpha$ ($\Sigma_{\rm SFR}^{\rm  H\alpha}$) and from FUV+24$\mu$m ($\Sigma_{\rm SFR}^{\rm FUV+24}$) generally agree in regions where $\Sigma_{\rm SFR}^{\rm H\alpha} > 10^{-2}~M_\odot~\rm yr^{-1}~kpc^{-2}$, with a scatter of about a factor of 2. However, in regions where $\Sigma_{\rm SFR}^{\rm H\alpha} < 10^{-2}~M_\odot~\rm yr^{-1}~kpc^{-2}$, $\Sigma_{\rm SFR}^{\rm FUV+24}$ is typically about twice as large as $\Sigma_{\rm SFR}^{\rm H\alpha}$. This difference increases as $\Sigma_{\rm SFR}^{\rm H\alpha}$ decreases. This trend, observed in many nearby galaxies \citep[e.g.,][]{leroy_phangsalma_2021}, is likely due to infrared cirrus components unrelated to active star formation, such as light from dust heated by older stars \citep[e.g.,][]{leroy_estimating_2012}. Consequently, in the bar region where star formation is less active, the mean $\Sigma_{\rm SFR}^{\rm FUV+24}$ is higher than $\Sigma_{\rm SFR}^{\rm H\alpha}$.

\begin{figure*}[]
 \begin{center}
  \includegraphics[width=180mm]{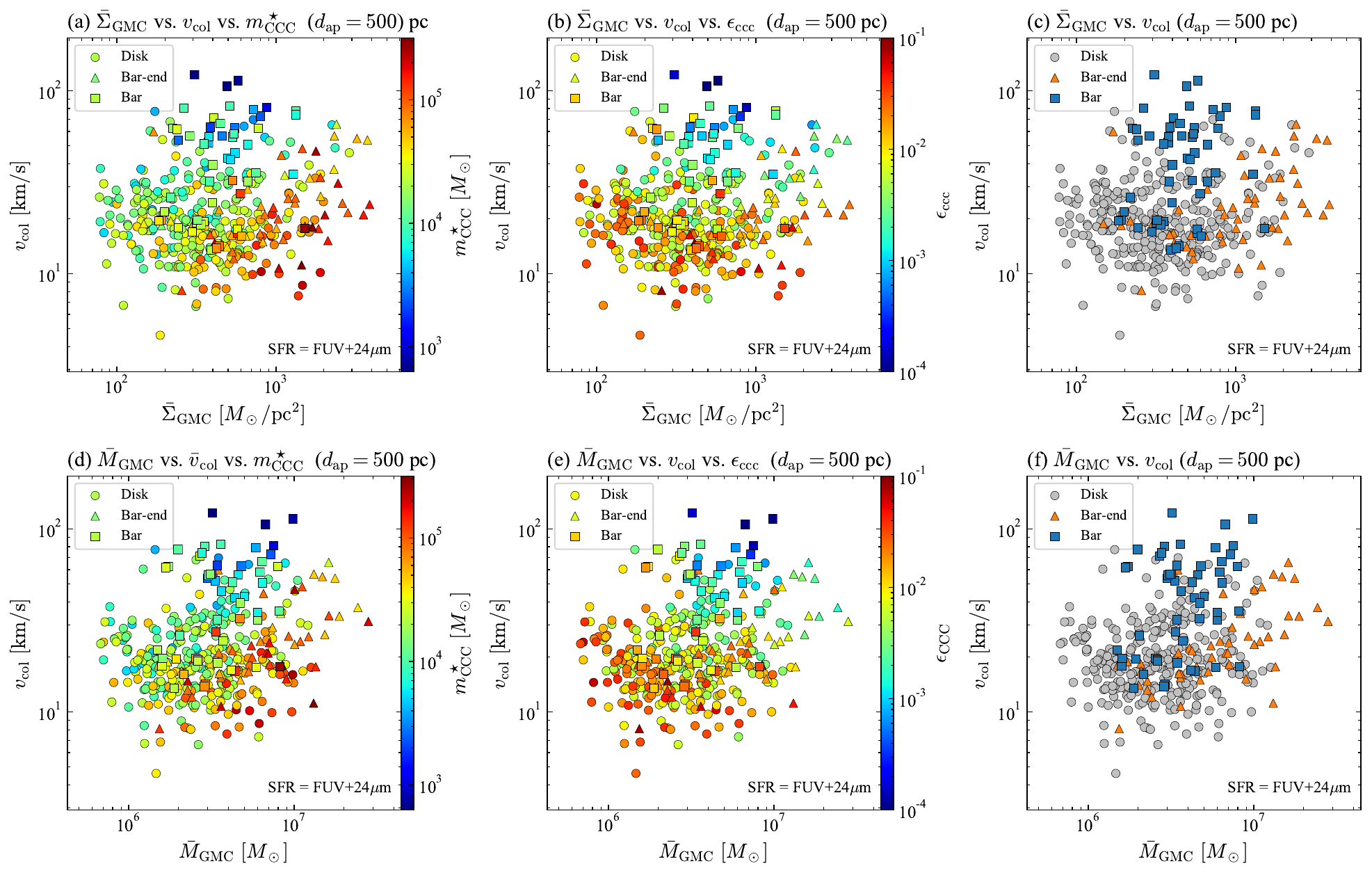} 
 \end{center}
\caption{ Same as Figure~\ref{fig:main_result}, but SFR is derived from FUV+24$\mu$m (Section~\ref{sec: SFR}). 
 }\label{fig:main_result_fuv24}
\end{figure*}

Figure \ref{fig:main_result_fuv24} is the same as Figure \ref{fig:main_result} but uses $\Sigma_{\rm SFR}^{\rm FUV+24}$.
Compared with $\Sigma_{\rm SFR}^{\rm H\alpha}$, the dependences on $\bar{\Sigma}_{\rm GMC}$ and $\bar{M}_{\rm GMC}$ are weaker.
This is because the $\Sigma_{\rm SFR}^{\rm H\alpha}$ tends to be low in regions where low-mass (or low-surface density) GMCs are present. Therefore, $\Sigma_{\rm SFR}^{\rm FUV+24}$ is higher than $\Sigma_{\rm SFR}^{\rm H\alpha}$ in these regions.
Consequently, the best-fit powers of $\bar{\Sigma}_{\rm GMC}$ and $\bar{M}_{\rm GMC}$ are smaller (Table~\ref{tab: fitting uncer}). 
Additionally, because the range of $\Sigma_{\rm SFR}^{\rm FUV+24}$ is narrower than that of $\Sigma_{\rm SFR}^{\rm H\alpha}$, the range of $m^\star_{\rm CCC}$ and $\epsilon_{\rm CCC}$ is also reduced, resulting in a shallower dependence on $v_{\rm col}$.

\subsubsection{CO(2--1)/CO(1--0) line ratio}\label{sec: CO(2--1)/CO(1--0) line ratio}

We assume a constant $R_{21}$ of 0.65 in calculating GMC mass based on kpc scale $R_{21}$ observations \citep{leroy_molecular_2013,den_brok_new_2021}.
However, $R_{21}$ on a kpc scale depends on galactic structures \citep[][]{yajima_R21_2021, Maeda_2023ApJ}: The $R_{21}$ in the bar-end region tends to be higher than that in the bar region. To account for the effect of this dependence, we recalculated our results using  $R_{21}$ map on a kpc scale of NGC~3627 obtained by \citet{Maeda_2023ApJ}. The low $R_{21}$ in the bar ($\sim0.4$) and high $R_{21}$ in the bar-end ($\sim0.8$) lessen the difference in GMC mass between these regions ($\sim0.3~\rm dex$; Table~\ref{tab: uncertainties}), thus weakening the dependence on $\bar{\Sigma}_{\rm GMC}$ and $\bar{M}_{\rm GMC}$ (Table~\ref{tab: fitting uncer}). However, it is uncertain whether this correction is accurate, because the $R_{21}$ measured on a kpc scale would differ from that on a GMC scale. In the bar region of NGC~1300, in particular, $R_{21}$ on the GMC scale is reported to be higher than the kpc-scale $R_{21}$ due to diffuse molecular gas, which is distributed on scales larger than a sub-kpc and would not contribute to the star formation activity \citep{Maeda2022_r21}. Therefore, the difference in $R_{21}$ between the bar and bar-end on a GMC scale is considered smaller than on a kpc scale. In fact, $R_{21}$ measurements at 200~pc scale in NGC~3627 show that difference in  $R_{21}$ between the bar ($\sim0.70$) and bar-end ($\sim0.86$) is as small as $\sim 0.1~\rm dex$  \citep{denbrok2023MNRAS.526.6347D}. This suggests that the dependences on $\bar{\Sigma}_{\rm GMC}$ and $\bar{M}_{\rm GMC}$ do not disappear. Accurate GMC mass measurements using CO(1--0) data on a GMC scale will be acquired in NGC 3627 as ALMA-FACTS project (Koda et al. in prep).
%
%Therefore, accurate GMC mass measurements require CO(1--0) data on a GMC scale, which will be acquired in NGC 3627 as ALMA-FACTS project (Koda et al. in prep).

\subsubsection{CO-to-H$_2$ conversion factor}\label{sec: conversion factor}
The choice of the $\alpha_{\rm CO}$ can be the largest source of uncertainty in measuring the GMC mass. As suggested by many studies \citep[e.g.,][]{Arimoto_xco_1996,Genzel_xco_2012,Accurso_xco_2017}, the $\alpha_{\rm CO}$ increases with the decrease in metallicity. Because the metallicity has a galactocentric radial gradient \citep[e.g.,][]{Sanchez_metallicity_2014}, using metallicity-dependent $\alpha_{\rm CO}$ may change our results.
The metallicity-dependent $\alpha_{\rm CO}$ is predicted by $\alpha_{\rm CO} = 4.35~{Z^\prime}^{-1.6}$ \citep{Accurso_xco_2017}, where $Z^\prime$ is the local gas phase abundance normalized to the solar value appropriate for the metallicity calibration by O3N2 method by \citet{PP04}. Using [OIII], [NII], H$\alpha$, and H$\beta$ images provided by PHANGS-MUSE projects \citep{Emsellem_phangs_muse2022}, we calculated the metallicity in NGC~3627 and adopted the local metallicity-dependent $\alpha_{\rm CO}$. As a result, the mean value is nearly solar metallicity. Furthermore, the galactocentric radial variation on the $\alpha_{\rm CO}$ is not seen in NGC~3627, which is consistent with other analysis \citep[e.g.,][]{Kreckel_2019ApJ}.  Consequently, our results do not change.

The $\alpha_{\rm CO}$ can be estimated using dust, CO, and H$_{\rm I}$ data under the assumption that the dust-to-gas ratio is constant at a kpc scale \citep{Sandstrom_xco_2013}, which is a method independent of the metallicity-based method described above. While the metallicity-based $\alpha_{\rm CO}$ in NGC~3627 is not significantly different from the Milky Way $\alpha_{\rm CO}$, the dust-based $\alpha_{\rm CO}$ are much lower, ranging from 0.51 to $1.2~M_\odot~\rm (K~km~s^{-1}~pc^{-2})^{-1}$ \citep{Sandstrom_xco_2013,Yasuda2023PASJ...75..743Y}. This suggests that the $M_{\rm GMC}$ and $\Sigma_{\rm GMC}$ could be 4–9 times smaller.
Furthermore, unlike metallicity-based $\alpha_{\rm CO}$, dust-based $\alpha_{\rm CO}$ was reported to exhibit a radial gradient, which could influence the results related to the dependences on $\bar{\Sigma}_{\rm GMC}$ and $\bar{M}_{\rm GMC}$. The $\alpha_{\rm CO}$ varies significantly across different studies, highlighting the need for more precise estimations as a key challenge for future research.

\subsubsection{GMC identification}\label{sec: GMC identification}
In this study, we use the same parameters for GMC identification with PYCPROPS as those used in the PHANGS project \citep{Rosolowsky_GMC_2021MNRAS}, which are considered to be the most fiducial. However, here we assess the uncertainties in GMC identification. The parameter in PYCPROPS most likely to affect results is the threshold for $T_{\rm max}-T_{\rm merge}$ in local maxima identification. We apply a threshold of $T_{\rm max}-T_{\rm merge} > 2\sigma_{\rm rms}$. Tables \ref{tab: uncertainties} and \ref{tab: fitting uncer} show the results when this threshold is set to $3\sigma_{\rm rms}$. In this case, the number of identified GMCs decreases by 43\% to 1033, reducing the $n_{\rm GMC}$. However, our results do not change significantly. This stability results from applying the condition $N_{\rm CCC} \geq 0.1~\rm Myr^{-1}$. This condition selects high-speed CCCs. Furthermore, the $T_{\rm max}-T_{\rm merge}$ threshold does not affect the calculation of $v_{\rm col}$ in regions with high-speed CCC. The $T_{\rm max}-T_{\rm merge}$ threshold only determines whether two adjacent local maxima in PPV space (i.e., with similar line-of-sight velocities) are treated as a single maximum. Therefore, the threshold does not affect the calculation of $v_{\rm col}$ in regions with high-speed CCC because the GMCs are separated in the velocity space.

Spatial resolution also affects GMC identification. However, the results remain consistent when the CO(2--1) cube is convolved to 90~pc resolution (Tables \ref{tab: uncertainties} and \ref{tab: fitting uncer}).  The spatial resolution primarily influences the identification of spatially adjacent GMCs. Similar to the threshold of $T_{\rm max}-T_{\rm merge}$ discussed above, the spatial resolution does not affect $v_{\rm col}$ calculations when GMCs are well-separated in the velocity space.

\subsection{Comparison with sub-parsec scale studies}

Although $m^\star_{\rm CCC}$ and $\epsilon_{\rm CCC}$ inherently tend to decrease as $v_{\rm col}$ increases in this study (Section~\ref{sec: method}), recent sub-pc scale CCC simulations show the $m^\star_{\rm CCC}$ decreases with increasing $v_{\rm col}$ \citep[][]{takahira_cloud-cloud_2014,takahira_formation_2018,Sakre_2023MNRAS}.\citet{takahira_cloud-cloud_2014} and \citet{takahira_formation_2018} conducted sub-parsec resolution simulations of CCCs between two clouds with collision speeds ranging from 3 to 30~$\rm km~s^{-1}$, finding that higher collision velocities can shorten the gas accretion phase, thereby suppressing core growth and massive star formation. Consequently, the slope of the core mass function becomes steeper with increasing collision velocity, and the total mass of the cores formed decreases. Similar suppression effects are seen in magnetohydrodynamic simulations \citep{Sakre_2023MNRAS}.

\citet{takahira_formation_2018} further indicates that the number of massive cores tends to increase with the mass of the colliding cloud at a fixed collision velocity, which is qualitatively consistent with our results. There are two possible reasons for this increase. First, if the radius increases with mass, the collision duration becomes longer, allowing more time for gas to accrete onto the cores. Second, if the density increases with mass, the amount of surrounding gas available for core accretion also increases. The latter effect may be significant in NGC~3627, because no correlation is seen between $\bar{M}_{\rm GMC}$ and $\bar{R}_{\rm GMC}$ within the extracted apertures.

Given the uncertainties in our results (Section~\ref{sec: Uncertainties}), it remains unclear how $\epsilon_{\rm CCC}$ depends on mass or surface density. Previous studies proposed the CCCs are responsible for the correlation between efficiency and cloud mass \citep[e.g.,][]{scoville_high-mass_1986,ikuta_kinematical_1997,tan_star_2000}.
However, in the simulation by \citep{takahira_formation_2018}, increasing the mass or surface density tends to increase the number of high-mass cores, but it does not seem to increase $\epsilon_{\rm CCC}$.

Note that other simulations suggest star formation activity driven by CCC also depends on the angle between the magnetic field direction and the collision axis \citep[e.g.,][]{Inoue_Inutsuka_2009ApJ, Sakre_2021PASJ}. \citet{Sakre_2021PASJ} found that when the two axes are nearly parallel, the converging gas flow toward the collision front is stronger, promoting the formation of massive cores. This dependence is expected to be explored in future studies of CCCs in the Milky Way.

\subsection{Connection to galactic structures}
Our results support the scenario that variations in the collision velocity and mass (or density) of colliding GMCs across different galactic structures may explain the observed differences in SFE on a kpc scale.
%As described in Section~\ref{sec: dependence}, we find that apertures across different galactic structures follow common dependencies of $v_{\rm col}$ and $\bar{\Sigma}_{\rm GMC}$ ($\bar{M}_{\rm GMC}$) on $m^\star_{\rm CCC}$ and $\epsilon_{\rm CCC}$ (Figure~\ref{fig:fitting}). 
The different structures have different distributions in the parameter space of $v_{\rm col}$ and $\bar{\Sigma}_{\rm GMC}$ ($\bar{M}_{\rm GMC}$), with higher $\bar{\Sigma}_{\rm GMC}$ ($\bar{M}_{\rm GMC}$) in the bar-end regions and higher $v_{\rm col}$ in the bar regions compared to the disk region, leading to differences in star-forming activity across these structures.

The different distributions in the parameter space of $v_{\rm col}$ and $\bar{\Sigma}_{\rm GMC}$ ($\bar{M}_{\rm GMC}$) by structures are interpreted as follows.  The non-axisymmetric gravitational bar potential increases noncircular gas motion in the bar-end and bar region, and the crossing of cloud orbits is thought to be an environment in which molecular cloud collisions occur frequently. In the bar-end regions, gas accumulates not only because of the stagnation of the gas in the elongated elliptical orbit but also because of the inflow of the gas rotating in the disk \citep[e.g.,][]{downes_co_1996}, resulting in high $\bar{\Sigma}_{\rm GMC}$ ($\bar{M}_{\rm GMC}$) \citep[e.g.,][]{renaud_environmental_2015}. In the bar region,  the $v_{\rm col}$ becomes high due to the violent noncircular, streaming motions caused by the bar potential. Parsec-scale hydrodynamical simulations of barred galaxies found the collision velocity of the GMCs in the bar regions is larger than in the arm regions due to the bar potential \citep{fujimoto_giant_2014,fujimoto_environmental_2014,fujimoto_fast_2020}, which is consistent with our results. Furthermore, a part of the gas that fell into the center would have overshot and sprayed into the dust lane \citep[e.g.,][]{Sormani_2019MNRAS}, which could be one factor contributing to the increase in the $v_{\rm col}$.

\subsection{Comparison with previous observational studies}
\label{sec: comparison}
Previous CO observations on a kpc scale towards nearby barred galaxies showed that the velocity width in bar regions was larger than in arm and bar-end regions and that there was a negative correlation between velocity width and SFE on a kpc scale \citep[e.g.,][]{yajima_co_2019, Maeda_2023ApJ}. This result may reflect the negative correlation between the $\epsilon_{\rm CCC}$ and $v_{\rm col}$.

Compared to the CCC study in NGC~1300 \citep{Maeda_CCC_2021}, the correlation between collision velocity, GMC mass, and star formation activity is qualitatively consistent with our findings in NGC~3627. However, NGC~3627 exhibits larger values for GMC mass, collision velocity, and collision frequency than NGC~1300. The typical values in the bar of NGC~3627 are $10^{6.5}~M_\odot$, $44~\rm km~s^{-1}$, and $101~\rm Gyr^{-1}$, while in the bar of NGC~1300, they are $10^{5.5}M_\odot$, $20~\rm km~s^{-1}$, and $25~\rm Gyr^{-1}$. The higher collision velocity in NGC~3627 may be attributed to its stronger bar potential, as the stellar surface density derived from WISE 3.4$\mu$m data \citep{leroy_z_2019} is about five times higher in the bar of NGC~3627 ($\sim 600~M_\odot\rm pc^{-2}$) than that in NGC~1300 ($\sim 130~M_\odot~\rm pc^{-2}$). On a kpc scale, the gas surface density in NGC~3627 ($\sim 100~M_\odot~\rm pc^{-2}$) is also one order of magnitude higher than that in NGC~1300 ($\sim 10~M_\odot~\rm pc^{-2}$), leading to significantly larger GMC mass and surface density in NGC~3627. Due to the lower GMC number density and collision velocity in NGC~1300, the collision frequency is also lower. These comparisons suggest that CCC properties may vary depending on the characteristics of the host galaxy such as morphology, stellar mass, and gas mass. Investigating the correlation between CCC properties and these factors will be an important direction for future research.

\citet{Sun_2022AJ} estimated the GMC collision timescale ($t_{\rm CCC}$) for 80 PHANGS galaxies, reporting a typical value of 91 Myr—about three times longer than the 27 Myr observed in NGC 3627. Several factors could explain this discrepancy. First, since $t_{\rm CCC}$ in \citet{Sun_2022AJ} was calculated across all regions with molecular gas detection, it likely includes many regions where CCC is not the dominant star formation mechanism, potentially resulting in a larger $t_{\rm CCC}$.
Second, the difference could arise from the methods used to calculate the collision velocity. While our study assumes random motion, \citet{Sun_2022AJ} adopted the shear-induced collision model \citep{tan_star_2000}, which assumes CCCs occur only when clouds catch up with others in adjacent circular orbits due to orbital shear. This method may overestimate $t_{\rm CCC}$ since streaming motions and intersections were not considered.
Finally, NGC~3627 may be a unique disk galaxy among the 80 PHANGS galaxies, with a short collision timescale due to its high gas surface density and high GMC number density compared to other galaxies, as mentioned above. It has been suggested that a high gas surface density on a kpc scale is essential for CCC-driven star formation to dominate \citep[e.g.,][]{Komugi_2006PASJ...58..793K}.

\section{summary}

Focusing on the nearby barred galaxy NGC~3627, we quantitatively investigate the SFR and SFE of colliding GMCs, explore how GMC mass(surface density) and collision velocity are linked to galactic structures (disk, bar-end, and bar), and examine how these relationships impact the SFR and SFE of colliding GMCs.  Using ALMA CO(2–1) data at a 60~pc resolution, we identified GMCs with PYCPROPS and estimated the collision velocity ($v_{\rm col}$) in hexagonal apertures with a size of 500 pc, based on the line-of-sight velocities of GMCs and assuming random motion in a two-dimensional plane. Using the mean SFR in the apertures, derived from the attenuation-corrected H$\alpha$ image obtained by VLT MUSE, we calculated the total mass of stars formed per CCC ($m^\star_{\rm CCC}$) and the SFE per CCC ($\epsilon_{\rm CCC}$)  in the apertures where CCC-driven star formation is considered to dominate (i.e., $N_{\rm CCC} \geq 0.1~\rm Myr^{-1}$).
The main results obtained and conclusions are as follows.

\begin{enumerate}
   \item CCCs are estimated to be frequent in the bar, bar-end, and disk regions. In most of the regions with strong CO(2–1) emission, the number of collisions occurring per unit time ($N_{\rm CCC}$) exceeds $0.1~\rm Myr^{-1}$ (Figure~\ref{fig:hex_tile}(b)).
   \item The median $v_{\rm col}$ within apertures where $N_{\rm CCC} \geq 0.1~\rm Myr^{-1}$ is 19.4~$\rm kms^{-1}$. The $v_{\rm col}$ is higher in the bar (44.4~$\rm kms^{-1}$) compared to the bar-end (21.0~$\rm kms^{-1}$) and disk (18.4~$\rm km~s^{-1}$) regions  (Table~\ref{tab: physical prop 500pc}). Consequently, the median $\nu_{\rm CCC}$ is higher ($t_{\rm CCC}$ is shorter) in the bar (101.1~$\rm Gyr^{-1}$, 9.89~Myr) compared to the bar-end (46.1~$\rm Gyr^{-1}$, 21.7~Myr) and disk (36.5~$\rm Gyr^{-1}$, 27.3~Myr) regions. 
   
   \item The median $m^\star_{\rm CCC}$ and $\epsilon_{\rm CCC}$ are $10^{4.30}~M_\odot$ and 0.73\%, respectively. In the bar region, the median $m^\star_{\rm CCC}$ and $\epsilon_{\rm CCC}$ are lower, at $10^{3.84}~M_\odot$ and 0.18\%, while in the bar-end region, they are higher, at $10^{4.89}~M_\odot$ and 1.10\%, compared to $10^{4.28}~M_\odot$ and 0.75\% in the disk region (Table~\ref{tab: physical prop 500pc}). Both $m^\star_{\rm CCC}$ and $\epsilon_{\rm CCC}$ decrease along the bar from the bar-end toward the center (Figure~\ref{fig:vcol_mstar_eccc_map}).
   \item  The $m^\star_{\rm CCC}$ decreases with both increasing $v_{\rm col}$ and decreasing $\bar{\Sigma}_{\rm GMC}$ ($\bar{M}_{\rm GMC}$) and the $\epsilon_{\rm CCC}$ decreases with increasing $v_{\rm col}$ (Figures~\ref{fig:main_result}, \ref{fig:fitting}). Although these results include trends by definition, they suggest that faster CCCs shorten the gas accretion phase and low density(mass) limits the amount of gas available for accretion, thereby suppressing cloud core growth and massive star formation. 
   
   \item The different structures have different distributions in the parameter space of $v_{\rm col}$ and $\bar{\Sigma}_{\rm GMC}$ ($\bar{M}_{\rm GMC}$), with higher $\bar{\Sigma}_{\rm GMC}$ ($\bar{M}_{\rm GMC}$) in the bar-end regions and higher $v_{\rm col}$ in the bar regions compared to the disk region, leading to differences in star-forming activity across these structures.
   \item Although aperture size and the choice of SFR tracer contribute most to the uncertainties in our results, our results remain robust.
\end{enumerate}

In conclusion, our results support the scenario that variations in the collision velocity and mass(density) of colliding GMCs across different galactic structures explain the observed differences in SFE on a kpc scale.
However, as discussed in Section~\ref{sec: comparison}, CCC properties within galaxies may vary depending on the characteristics of the host galaxy, and more statistical surveys will be needed in the future. 
%The methods used in this study can be applied beyond disk galaxies. In particular, gas-rich colliding galaxies are thought to experience GMC collisions at higher velocities than those observed in disk galaxies \citep{Tsuge_2021PASJ...73S..35T,Tsuge_2021PASJ...73..417T}. The relationship between CCC properties and star formation in these violent environments is being investigated (Inoue et al. in prep). Similar to our study, star formation is likely dependent on both collision velocity and GMC mass, suggesting that fast collisions of massive GMCs play a crucial role in the formation of superstar clusters.

\newpage

\startlongtable
\begin{deluxetable*}{ccccccccccc}
\tablecaption{Physical properties within the apertures under different conditions.\label{tab: uncertainties}}
\tablewidth{0pt}
\tablehead{ 
   Region & \# &$\log \bar{M}_{\rm GMC}$ & $\log \bar{\Sigma}_{\rm GMC}$ & $\log \Sigma_{\rm SFR}^{\rm ap}$ & $n_{\rm GMC}$ & $v_{\rm col}$ & $\nu_{\rm CCC}$ & $N_{\rm CCC}$ & $\log  m^\star_{\rm CCC} $  & $\epsilon_{\rm CCC}$\\ 
      & & $M_\odot$ & $M_\odot~\rm pc^{-2}$ & $M_\odot~\rm yr^{-1}~kpc^{-2}$& $\rm kpc^{-2}$ & $\rm km~s^{-1}$ &$\rm Gyr^{-1}$ & $\rm Myr^{-1}$ & $M_\odot$ & \% 
 }
\decimalcolnumbers
\startdata
      \multicolumn{11}{c}{$d_{\rm ap}$ = 300~pc (Section~\ref{sec: Aperture size})}\\
All & 411 & $6.47_{-0.21}^{+0.25}$ & $2.56_{-0.25}^{+0.32}$ & $-1.74_{-0.37}^{+0.44}$ & $27.7_{-0.0}^{+9.2}$ & $23.0_{-6.4}^{+11.0}$ & $69.1_{-18.7}^{+38.6}$ & $0.22_{-0.08}^{+0.21}$ & $3.90_{-0.47}^{+0.51}$ & $0.28_{-0.19}^{+0.47}$ \\
Disk & 288 & $6.40_{-0.19}^{+0.24}$ & $2.49_{-0.24}^{+0.34}$ & $-1.76_{-0.41}^{+0.41}$ & $27.7_{-0.0}^{+9.2}$ & $21.2_{-5.5}^{+8.2}$ & $64.0_{-15.7}^{+25.6}$ & $0.19_{-0.06}^{+0.12}$ & $3.89_{-0.43}^{+0.46}$ & $0.32_{-0.20}^{+0.44}$ \\
Bar-end & 56 & $6.80_{-0.25}^{+0.26}$ & $2.96_{-0.19}^{+0.27}$ & $-1.09_{-0.47}^{+0.41}$ & $37.0_{-9.2}^{+9.2}$ & $25.0_{-6.8}^{+14.4}$ & $74.0_{-25.3}^{+56.8}$ & $0.26_{-0.11}^{+0.28}$ & $4.62_{-0.57}^{+0.22}$ & $0.55_{-0.36}^{+0.74}$ \\
Bar & 67 & $6.55_{-0.13}^{+0.15}$ & $2.58_{-0.10}^{+0.28}$ & $-1.97_{-0.17}^{+0.36}$ & $27.7_{-0.0}^{+9.2}$ & $43.2_{-18.9}^{+16.4}$ & $148.2_{-66.6}^{+92.8}$ & $0.44_{-0.21}^{+0.52}$ & $3.41_{-0.31}^{+0.47}$ & $0.07_{-0.04}^{+0.16}$ \\
      \cline{1-11}
    \multicolumn{11}{c}{$d_{\rm ap}$ = 400~pc (Section~\ref{sec: Aperture size})}\\
All & 423 & $6.49_{-0.20}^{+0.23}$ & $2.62_{-0.27}^{+0.27}$ & $-1.69_{-0.44}^{+0.43}$ & $26.0_{-5.2}^{+5.2}$ & $20.9_{-5.1}^{+10.0}$ & $48.7_{-12.9}^{+26.1}$ & $0.21_{-0.07}^{+0.17}$ & $4.14_{-0.38}^{+0.54}$ & $0.52_{-0.32}^{+0.68}$ \\
Disk & 295 & $6.41_{-0.20}^{+0.22}$ & $2.55_{-0.27}^{+0.28}$ & $-1.72_{-0.44}^{+0.37}$ & $26.0_{-5.2}^{+5.2}$ & $19.6_{-4.6}^{+6.9}$ & $46.1_{-11.7}^{+18.0}$ & $0.20_{-0.06}^{+0.12}$ & $4.12_{-0.33}^{+0.45}$ & $0.55_{-0.31}^{+0.57}$ \\
Bar-end & 62 & $6.76_{-0.18}^{+0.28}$ & $3.05_{-0.33}^{+0.25}$ & $-0.98_{-0.34}^{+0.27}$ & $26.0_{-5.2}^{+5.2}$ & $20.4_{-3.7}^{+10.1}$ & $50.9_{-14.4}^{+25.2}$ & $0.23_{-0.06}^{+0.15}$ & $4.94_{-0.60}^{+0.25}$ & $1.17_{-0.81}^{+1.75}$ \\
Bar & 66 & $6.55_{-0.15}^{+0.15}$ & $2.61_{-0.14}^{+0.23}$ & $-2.01_{-0.15}^{+0.44}$ & $23.4_{-2.6}^{+7.8}$ & $38.6_{-16.9}^{+19.7}$ & $105.1_{-58.5}^{+53.6}$ & $0.45_{-0.28}^{+0.39}$ & $3.76_{-0.42}^{+0.36}$ & $0.16_{-0.10}^{+0.28}$ \\
      \cline{1-11}
    \multicolumn{11}{c}{$d_{\rm ap}$ = 500~pc (Same as Table~\ref{tab: physical prop 500pc}: Section~\ref{sec: results})}\\
All & 406 & $6.49_{-0.22}^{+0.18}$ & $2.59_{-0.24}^{+0.25}$ & $-1.77_{-0.37}^{+0.42}$ & $20.0_{-3.3}^{+5.8}$ & $19.4_{-4.5}^{+10.7}$ & $38.1_{-10.1}^{+21.0}$ & $0.22_{-0.07}^{+0.21}$ & $4.30_{-0.44}^{+0.45}$ & $0.73_{-0.44}^{+0.88}$ \\
Disk & 301 & $6.42_{-0.21}^{+0.18}$ & $2.52_{-0.24}^{+0.25}$ & $-1.83_{-0.36}^{+0.38}$ & $20.0_{-3.3}^{+3.3}$ & $18.4_{-4.5}^{+7.6}$ & $36.5_{-9.6}^{+15.6}$ & $0.20_{-0.05}^{+0.13}$ & $4.28_{-0.42}^{+0.37}$ & $0.75_{-0.43}^{+0.86}$ \\
Bar-end & 51 & $6.81_{-0.27}^{+0.23}$ & $3.05_{-0.32}^{+0.23}$ & $-1.04_{-0.32}^{+0.28}$ & $23.3_{-3.3}^{+3.3}$ & $21.0_{-4.5}^{+9.1}$ & $46.1_{-12.1}^{+16.4}$ & $0.31_{-0.09}^{+0.13}$ & $4.89_{-0.32}^{+0.23}$ & $1.10_{-0.50}^{+1.97}$ \\
Bar & 54 & $6.54_{-0.10}^{+0.21}$ & $2.65_{-0.14}^{+0.11}$ & $-1.89_{-0.26}^{+0.38}$ & $23.3_{-6.7}^{+3.3}$ & $44.4_{-24.7}^{+18.3}$ & $101.1_{-64.3}^{+62.3}$ & $0.59_{-0.38}^{+0.64}$ & $3.84_{-0.33}^{+0.38}$ & $0.18_{-0.11}^{+0.39}$ \\
      \cline{1-11}
    \multicolumn{11}{c}{$d_{\rm ap}$ = 600~pc (Section~\ref{sec: Aperture size})}\\
All & 372 & $6.48_{-0.21}^{+0.20}$ & $2.58_{-0.25}^{+0.20}$ & $-1.83_{-0.36}^{+0.39}$ & $18.5_{-4.6}^{+4.6}$ & $19.6_{-4.2}^{+10.5}$ & $36.3_{-10.8}^{+17.6}$ & $0.26_{-0.10}^{+0.20}$ & $4.35_{-0.37}^{+0.38}$ & $0.78_{-0.38}^{+0.76}$ \\
Disk & 284 & $6.42_{-0.19}^{+0.21}$ & $2.49_{-0.22}^{+0.22}$ & $-1.90_{-0.33}^{+0.37}$ & $16.2_{-2.3}^{+4.6}$ & $18.9_{-4.1}^{+6.0}$ & $32.6_{-8.0}^{+12.0}$ & $0.22_{-0.07}^{+0.16}$ & $4.35_{-0.33}^{+0.33}$ & $0.88_{-0.42}^{+0.66}$ \\
Bar-end & 43 & $6.75_{-0.31}^{+0.24}$ & $3.04_{-0.37}^{+0.24}$ & $-1.03_{-0.49}^{+0.28}$ & $23.1_{-4.6}^{+4.6}$ & $21.2_{-4.8}^{+12.7}$ & $44.9_{-12.3}^{+22.8}$ & $0.44_{-0.16}^{+0.23}$ & $4.88_{-0.37}^{+0.23}$ & $1.32_{-0.78}^{+1.62}$ \\
Bar & 45 & $6.57_{-0.19}^{+0.12}$ & $2.65_{-0.14}^{+0.16}$ & $-1.85_{-0.34}^{+0.33}$ & $18.5_{-2.3}^{+6.9}$ & $51.8_{-25.6}^{+14.0}$ & $79.0_{-44.4}^{+87.3}$ & $0.65_{-0.38}^{+1.21}$ & $3.96_{-0.25}^{+0.23}$ & $0.20_{-0.08}^{+0.46}$ \\
    \cline{1-11}
    \multicolumn{11}{c}{$d_{\rm ap}$ = 700~pc (Section~\ref{sec: Aperture size})}\\
All & 326 & $6.47_{-0.22}^{+0.19}$ & $2.54_{-0.23}^{+0.24}$ & $-1.85_{-0.34}^{+0.39}$ & $15.3_{-3.4}^{+5.1}$ & $20.7_{-4.8}^{+7.9}$ & $32.4_{-9.1}^{+15.5}$ & $0.27_{-0.10}^{+0.27}$ & $4.43_{-0.38}^{+0.33}$ & $0.96_{-0.53}^{+0.76}$ \\
Disk & 247 & $6.43_{-0.22}^{+0.20}$ & $2.46_{-0.19}^{+0.26}$ & $-1.92_{-0.34}^{+0.34}$ & $15.3_{-3.4}^{+3.4}$ & $19.9_{-4.6}^{+5.3}$ & $30.4_{-7.5}^{+10.6}$ & $0.25_{-0.08}^{+0.19}$ & $4.41_{-0.36}^{+0.29}$ & $0.97_{-0.43}^{+0.73}$ \\
Bar-end & 39 & $6.73_{-0.32}^{+0.31}$ & $3.02_{-0.41}^{+0.23}$ & $-1.07_{-0.45}^{+0.23}$ & $18.7_{-3.4}^{+3.4}$ & $21.4_{-4.2}^{+12.7}$ & $44.3_{-15.7}^{+23.5}$ & $0.50_{-0.24}^{+0.41}$ & $4.87_{-0.28}^{+0.24}$ & $1.88_{-1.20}^{+1.04}$ \\
Bar & 40 & $6.54_{-0.17}^{+0.13}$ & $2.63_{-0.15}^{+0.11}$ & $-1.82_{-0.37}^{+0.23}$ & $17.0_{-5.1}^{+6.8}$ & $46.9_{-27.9}^{+15.1}$ & $69.9_{-41.3}^{+76.6}$ & $0.67_{-0.45}^{+1.41}$ & $4.04_{-0.19}^{+0.44}$ & $0.39_{-0.23}^{+0.59}$ \\
    \cline{1-11}
    \multicolumn{11}{c}{$t_{\rm CCC} \leq 30~\rm Myr$ (Section~\ref{sec: Aperture extraction criteria})}\\
All & 264 & $6.52_{-0.22}^{+0.20}$ & $2.59_{-0.27}^{+0.29}$ & $-1.79_{-0.34}^{+0.45}$ & $23.3_{-6.7}^{+3.3}$ & $25.8_{-6.5}^{+10.0}$ & $52.2_{-13.7}^{+30.9}$ & $0.32_{-0.11}^{+0.27}$ & $4.15_{-0.40}^{+0.40}$ & $0.48_{-0.28}^{+0.59}$ \\
Disk & 181 & $6.43_{-0.21}^{+0.18}$ & $2.45_{-0.23}^{+0.33}$ & $-1.85_{-0.36}^{+0.42}$ & $20.0_{-3.3}^{+6.7}$ & $24.5_{-5.8}^{+6.7}$ & $46.6_{-9.0}^{+18.3}$ & $0.28_{-0.08}^{+0.20}$ & $4.16_{-0.37}^{+0.25}$ & $0.52_{-0.28}^{+0.58}$ \\
Bar-end & 39 & $6.85_{-0.23}^{+0.25}$ & $3.08_{-0.22}^{+0.25}$ & $-1.04_{-0.35}^{+0.35}$ & $23.3_{-1.7}^{+6.7}$ & $23.9_{-5.4}^{+10.2}$ & $53.7_{-14.3}^{+18.5}$ & $0.38_{-0.10}^{+0.19}$ & $4.75_{-0.30}^{+0.34}$ & $0.99_{-0.63}^{+0.33}$ \\
Bar & 44 & $6.56_{-0.08}^{+0.21}$ & $2.68_{-0.13}^{+0.09}$ & $-1.89_{-0.26}^{+0.35}$ & $23.3_{-3.3}^{+6.7}$ & $54.9_{-22.5}^{+12.7}$ & $128.5_{-53.6}^{+63.0}$ & $0.93_{-0.59}^{+0.46}$ & $3.69_{-0.22}^{+0.35}$ & $0.13_{-0.08}^{+0.33}$ \\
    \cline{1-11}
\multicolumn{11}{c}{SFR = FUV +24$\mu$m (Section~\ref{sec: SFR})}\\
All & 406 & $6.49_{-0.22}^{+0.18}$ & $2.59_{-0.24}^{+0.25}$ & $-1.71_{-0.18}^{+0.30}$ & $20.0_{-3.3}^{+5.8}$ & $19.4_{-4.5}^{+10.7}$ & $38.1_{-10.1}^{+21.0}$ & $0.22_{-0.07}^{+0.21}$ & $4.40_{-0.30}^{+0.27}$ & $0.86_{-0.39}^{+0.79}$ \\
Disk & 301 & $6.42_{-0.21}^{+0.18}$ & $2.52_{-0.24}^{+0.25}$ & $-1.77_{-0.22}^{+0.26}$ & $20.0_{-3.3}^{+3.3}$ & $18.4_{-4.5}^{+7.6}$ & $36.5_{-9.6}^{+15.6}$ & $0.20_{-0.05}^{+0.13}$ & $4.37_{-0.25}^{+0.22}$ & $0.85_{-0.35}^{+0.65}$ \\
Bar-end & 51 & $6.81_{-0.27}^{+0.23}$ & $3.05_{-0.32}^{+0.23}$ & $-1.03_{-0.24}^{+0.22}$ & $23.3_{-3.3}^{+3.3}$ & $21.0_{-4.5}^{+9.1}$ & $46.1_{-12.1}^{+16.4}$ & $0.31_{-0.09}^{+0.13}$ & $4.92_{-0.22}^{+0.16}$ & $1.52_{-0.82}^{+0.89}$ \\
Bar & 54 & $6.54_{-0.10}^{+0.21}$ & $2.65_{-0.14}^{+0.11}$ & $-1.72_{-0.06}^{+0.20}$ & $23.3_{-6.7}^{+3.3}$ & $44.4_{-24.7}^{+18.3}$ & $101.1_{-64.3}^{+62.3}$ & $0.59_{-0.38}^{+0.64}$ & $4.05_{-0.29}^{+0.40}$ & $0.33_{-0.21}^{+0.73}$ \\
    \cline{1-11}
    \multicolumn{11}{c}{variable $R_{21}$ (Section~\ref{sec: CO(2--1)/CO(1--0) line ratio})}\\
All & 406 & $6.60_{-0.20}^{+0.19}$ & $2.73_{-0.23}^{+0.19}$ & $-1.77_{-0.37}^{+0.42}$ & $20.0_{-3.3}^{+5.8}$ & $19.4_{-4.5}^{+10.7}$ & $38.1_{-10.1}^{+21.0}$ & $0.22_{-0.07}^{+0.21}$ & $4.30_{-0.44}^{+0.45}$ & $0.58_{-0.38}^{+0.84}$ \\
Disk & 301 & $6.58_{-0.21}^{+0.18}$ & $2.71_{-0.24}^{+0.17}$ & $-1.83_{-0.36}^{+0.38}$ & $20.0_{-3.3}^{+3.3}$ & $18.4_{-4.5}^{+7.6}$ & $36.5_{-9.6}^{+15.6}$ & $0.20_{-0.05}^{+0.13}$ & $4.28_{-0.42}^{+0.37}$ & $0.58_{-0.35}^{+0.64}$ \\
Bar-end & 51 & $6.62_{-0.14}^{+0.34}$ & $2.96_{-0.26}^{+0.13}$ & $-1.04_{-0.32}^{+0.28}$ & $23.3_{-3.3}^{+3.3}$ & $21.0_{-4.5}^{+9.1}$ & $46.1_{-12.1}^{+16.4}$ & $0.31_{-0.09}^{+0.13}$ & $4.89_{-0.32}^{+0.23}$ & $1.81_{-1.14}^{+1.92}$ \\
Bar & 54 & $6.69_{-0.20}^{+0.22}$ & $2.78_{-0.20}^{+0.22}$ & $-1.89_{-0.26}^{+0.38}$ & $23.3_{-6.7}^{+3.3}$ & $44.4_{-24.7}^{+18.3}$ & $101.1_{-64.3}^{+62.3}$ & $0.59_{-0.38}^{+0.64}$ & $3.84_{-0.33}^{+0.38}$ & $0.15_{-0.10}^{+0.35}$ \\
    \cline{1-11}
    \multicolumn{11}{c}{metallicity-dependent $\alpha_{\rm CO}$ (Section~\ref{sec: conversion factor})}\\
All & 406 & $6.49_{-0.22}^{+0.18}$ & $2.59_{-0.24}^{+0.24}$ & $-1.77_{-0.37}^{+0.42}$ & $20.0_{-3.3}^{+5.8}$ & $19.4_{-4.5}^{+10.7}$ & $38.1_{-10.1}^{+21.0}$ & $0.22_{-0.07}^{+0.21}$ & $4.30_{-0.44}^{+0.45}$ & $0.71_{-0.44}^{+0.87}$ \\
Disk & 301 & $6.43_{-0.21}^{+0.19}$ & $2.52_{-0.24}^{+0.25}$ & $-1.83_{-0.36}^{+0.38}$ & $20.0_{-3.3}^{+3.3}$ & $18.4_{-4.5}^{+7.6}$ & $36.5_{-9.6}^{+15.6}$ & $0.20_{-0.05}^{+0.13}$ & $4.28_{-0.42}^{+0.37}$ & $0.75_{-0.44}^{+0.84}$ \\
Bar-end & 51 & $6.82_{-0.27}^{+0.23}$ & $3.04_{-0.31}^{+0.23}$ & $-1.04_{-0.32}^{+0.28}$ & $23.3_{-3.3}^{+3.3}$ & $21.0_{-4.5}^{+9.1}$ & $46.1_{-12.1}^{+16.4}$ & $0.31_{-0.09}^{+0.13}$ & $4.89_{-0.32}^{+0.23}$ & $1.10_{-0.51}^{+1.95}$ \\
Bar & 54 & $6.55_{-0.11}^{+0.21}$ & $2.65_{-0.13}^{+0.12}$ & $-1.89_{-0.26}^{+0.38}$ & $23.3_{-6.7}^{+3.3}$ & $44.4_{-24.7}^{+18.3}$ & $101.1_{-64.3}^{+62.3}$ & $0.59_{-0.38}^{+0.64}$ & $3.84_{-0.33}^{+0.38}$ & $0.18_{-0.11}^{+0.38}$ \\
    \cline{1-11}
    \multicolumn{11}{c}{$T_{\rm max}-T_{\rm merge} > 3\sigma_{\rm rms}$  (Section~\ref{sec: GMC identification}) }\\
All & 186 & $6.78_{-0.26}^{+0.23}$ & $2.73_{-0.33}^{+0.25}$ & $-1.60_{-0.43}^{+0.39}$ & $16.6_{-3.3}^{+3.3}$ & $26.3_{-8.4}^{+11.9}$ & $45.7_{-11.9}^{+29.7}$ & $0.19_{-0.05}^{+0.15}$ & $4.53_{-0.41}^{+0.42}$ & $0.68_{-0.44}^{+0.74}$ \\
Disk & 120 & $6.70_{-0.25}^{+0.20}$ & $2.57_{-0.31}^{+0.26}$ & $-1.64_{-0.40}^{+0.28}$ & $13.3_{-0.0}^{+3.3}$ & $25.0_{-7.6}^{+6.9}$ & $42.0_{-8.2}^{+17.3}$ & $0.18_{-0.05}^{+0.14}$ & $4.44_{-0.32}^{+0.41}$ & $0.73_{-0.46}^{+0.68}$ \\
Bar-end & 33 & $7.05_{-0.23}^{+0.24}$ & $3.13_{-0.26}^{+0.13}$ & $-0.93_{-0.14}^{+0.26}$ & $16.6_{-0.0}^{+3.3}$ & $23.7_{-7.1}^{+10.6}$ & $42.0_{-11.5}^{+21.4}$ & $0.21_{-0.05}^{+0.08}$ & $5.24_{-0.29}^{+0.10}$ & $1.24_{-0.58}^{+2.08}$ \\
Bar & 33 & $6.90_{-0.23}^{+0.15}$ & $2.73_{-0.11}^{+0.12}$ & $-2.01_{-0.11}^{+0.36}$ & $13.3_{-3.3}^{+3.3}$ & $50.8_{-12.4}^{+20.1}$ & $84.1_{-26.5}^{+58.2}$ & $0.32_{-0.15}^{+0.34}$ & $3.93_{-0.22}^{+0.39}$ & $0.14_{-0.08}^{+0.19}$ \\
    \cline{1-11}
    \multicolumn{11}{c}{90~pc resolution (Section~\ref{sec: GMC identification}) }\\
All & 261 & $6.61_{-0.24}^{+0.20}$ & $2.44_{-0.32}^{+0.36}$ & $-1.89_{-0.32}^{+0.39}$ & $13.3_{-3.3}^{+3.3}$ & $30.3_{-8.0}^{+14.0}$ & $57.8_{-18.6}^{+40.0}$ & $0.20_{-0.06}^{+0.22}$ & $4.19_{-0.39}^{+0.40}$ & $0.43_{-0.26}^{+0.67}$ \\
Disk & 195 & $6.49_{-0.18}^{+0.25}$ & $2.32_{-0.22}^{+0.37}$ & $-1.92_{-0.33}^{+0.34}$ & $13.3_{-3.3}^{+3.3}$ & $29.0_{-7.6}^{+9.0}$ & $49.6_{-10.6}^{+36.6}$ & $0.19_{-0.05}^{+0.18}$ & $4.19_{-0.34}^{+0.34}$ & $0.46_{-0.25}^{+0.64}$ \\
Bar-end & 20 & $7.22_{-0.34}^{+0.28}$ & $3.18_{-0.49}^{+0.11}$ & $-1.03_{-0.19}^{+0.43}$ & $13.3_{-0.8}^{+3.3}$ & $26.3_{-5.0}^{+11.3}$ & $43.4_{-9.5}^{+27.5}$ & $0.18_{-0.06}^{+0.10}$ & $5.17_{-0.17}^{+0.31}$ & $1.20_{-0.32}^{+0.80}$ \\
Bar & 46 & $6.75_{-0.19}^{+0.17}$ & $2.58_{-0.28}^{+0.22}$ & $-2.00_{-0.15}^{+0.27}$ & $13.3_{-3.3}^{+5.8}$ & $52.9_{-18.6}^{+11.1}$ & $110.6_{-49.1}^{+43.9}$ & $0.42_{-0.22}^{+0.42}$ & $3.93_{-0.37}^{+0.33}$ & $0.14_{-0.09}^{+0.22}$ \\
\enddata
\tablecomments{
Each physical property is noted as $M^{+D75}_{-D25}$, where $M$, $D25$, and $D75$ are the median, the distance to the 25th percentile from the median, and the distance to the 75th percentile from the median of the number distribution, respectively.}
\end{deluxetable*}

\newpage

\begin{longrotatetable}
    
\begin{deluxetable*}{lccc|ccc|ccc}
\tablecaption{Fitting results of the dependence of $\bar{m}_{\rm CCC}^\star$ and $\epsilon_{\rm CCC}$ on $v_{\rm col}$ and $\bar{\Sigma}_{\rm GMC}$.\label{tab: fitting uncer}}
\tablewidth{0pt}
\tablehead{ 
    &\multicolumn{3}{c}{$\log{\bar{m}_{\rm CCC}^\star} = a \log{v_{\rm col}}+ b\log{\bar{\Sigma}_{\rm GMC}} +c$}
    &\multicolumn{3}{c}{$\log{\epsilon_{\rm CCC}} =  d \log{v_{\rm col}}+ e\log{\bar{\Sigma}_{\rm GMC}} +f$}
    &\multicolumn{3}{c}{$\log{\bar{m}_{\rm CCC}^\star} = g \log{v_{\rm col}}+ h\log{\bar{M}_{\rm GMC}} +i$} \\
    & $a$ & $b$ & $c$ & $d$ & $e$ & $f$ & $g$ & $h$ & $i$
 }
\decimalcolnumbers
\startdata
$d_{\rm ap} = 300$~pc &$-1.29 \pm 0.10$  & $0.98 \pm 0.06$ & $3.15 \pm 0.20$ & $-1.45 \pm 0.11$ & $0.30 \pm 0.06$ & $-1.35 \pm 0.22$ & $-1.39 \pm 0.11$  & $0.92 \pm 0.08$ & $-0.12 \pm 0.51$ \\
$d_{\rm ap} = 400$~pc &$-1.39 \pm 0.09$  & $1.03 \pm 0.05$ & $3.34 \pm 0.18$ & $-1.59 \pm 0.10$ & $0.35 \pm 0.06$ & $-1.08 \pm 0.20$ & $-1.57 \pm 0.11$  & $1.04 \pm 0.07$ & $-0.47 \pm 0.47$ \\
$d_{\rm ap} = 500$~pc & $-1.31 \pm 0.08$  & $1.00 \pm 0.05$ & $3.43 \pm 0.17$ & $-1.47 \pm 0.09$ & $0.31 \pm 0.06$ & $-1.04 \pm 0.18$ & $-1.41 \pm 0.09$  & $1.02 \pm 0.07$ & $-0.47 \pm 0.46$ \\
$d_{\rm ap} = 600$~pc & $-1.32 \pm 0.09$  & $0.94 \pm 0.06$ & $3.68 \pm 0.17$ & $-1.44 \pm 0.09$ & $0.24 \pm 0.06$ & $-0.84 \pm 0.18$ & $-1.35 \pm 0.09$  & $0.99 \pm 0.07$ & $-0.27 \pm 0.46$ \\
$d_{\rm ap} = 700$~pc & $-1.36 \pm 0.09$  & $0.89 \pm 0.06$ & $3.97 \pm 0.18$ & $-1.43 \pm 0.10$ & $0.18 \pm 0.06$ & $-0.59 \pm 0.18$ & $-1.34 \pm 0.10$  & $0.94 \pm 0.08$ & $0.13 \pm 0.47$ \\
$t_{\rm CCC} \leq 30~\rm Myr$ & $-1.45 \pm 0.12$  & $0.90 \pm 0.06$ & $3.88 \pm 0.21$ & $-1.60 \pm 0.13$ & $0.23 \pm 0.07$ & $-0.65 \pm 0.24$ & $-1.48 \pm 0.14$  & $0.93 \pm 0.09$ & $0.23 \pm 0.55$ \\
SFR = FUV +24$\mu$m & $-0.99 \pm 0.07$  & $0.58 \pm 0.05$ & $4.20 \pm 0.14$ & $-1.15 \pm 0.08$ & $-0.11 \pm 0.05$ & $-0.27 \pm 0.16$ & $-1.05 \pm 0.08$  & $0.61 \pm 0.06$ & $1.82 \pm 0.37$ \\
variable $R_{21}$ & $-1.28 \pm 0.10$  & $0.86 \pm 0.08$ & $3.65 \pm 0.23$ & $-1.44 \pm 0.11$ & $0.17 \pm 0.08$ & $-0.83 \pm 0.25$ & $-1.27 \pm 0.11$  & $0.57 \pm 0.09$ & $2.24 \pm 0.59$ \\
metallicity-dependent $\alpha_{\rm CO}$ & $-1.31 \pm 0.08$  & $1.00 \pm 0.06$ & $3.43 \pm 0.17$ & $-1.47 \pm 0.09$ & $0.31 \pm 0.06$ & $-1.04 \pm 0.18$ & $-1.41 \pm 0.09$  & $1.02 \pm 0.07$ & $-0.43 \pm 0.47$ \\
$T_{\rm max}-T_{\rm merge} > 3\sigma_{\rm rms}$ & $-1.41 \pm 0.12$  & $0.89 \pm 0.07$ & $4.15 \pm 0.24$ & $-1.63 \pm 0.13$ & $0.15 \pm 0.08$ & $-0.33 \pm 0.27$ & $-1.52 \pm 0.14$  & $0.85 \pm 0.10$ & $0.91 \pm 0.63$ \\
90~pc resolution & $-1.36 \pm 0.13$  & $0.78 \pm 0.06$ & $4.35 \pm 0.23$ & $-1.49 \pm 0.13$ & $0.12 \pm 0.06$ & $-0.46 \pm 0.24$ & $-1.46 \pm 0.13$  & $0.98 \pm 0.08$ & $-0.07 \pm 0.56$ \\
\enddata
\tablecomments{In the fitting of $\log{\epsilon_{\rm CCC}} = j \log{v_{\rm col}} + k \log{\bar{M}_{\rm GMC}} + l$, the best-fitting coefficients are given by $g = j$, $h = k + 1$, and $i = l$ due to the relation $\epsilon_{\rm CCC} = \bar{m}_{\rm CCC}^\star / \bar{M}_{\rm GMC}$. }
\end{deluxetable*}

\end{longrotatetable}

\begin{acknowledgments}
We sincerely thank the anonymous reviewer for the insightful comments.
F.M., K.O., F.E., and Y.F., are supported by JSPS KAKENHI grant No. JP23K13142, JP23K03458, JP20H00172, and JP22K20387, respectively.
F.M. was supported by the ALMA Japan Research Grant of NAOJ ALMA Project, NAOJ-ALMA-351.
This paper makes use of the following ALMA data:
ADS/JAO.ALMA \#2015.1.00956.S.
%Part of these projects have been processed as the PHANGS-ALMA CO (2--1) survey.
ALMA is a partnership of ESO (representing its member states), NSF (USA), and NINS (Japan), together with NRC (Canada), MOST and ASIAA (Taiwan), and KASI (Republic of Korea), in cooperation with the Republic of Chile. The Joint ALMA Observatory is operated by ESO, AUI/NRAO, and NAOJ. 
Data analysis was in part carried out on the Multi-wavelength Data Analysis System operated by the Astronomy Data Center (ADC), NAOJ.
This research also made use of APLpy, an open-source plotting package for Python \citep{Robitaille_2012}.

\end{acknowledgments}


\vspace{5mm}
\facilities{ALMA, VLT, GALEX, Spitzer}


\software{CASA \citep{mcmullin_casa_ASPCS}, Astropy \citep{astropy_2018}, APLpy \citep{Robitaille_2012}}, NumPy \citep{Numpy_harris2020array}, SciPy \citep{Virtanen_scipy}, astroquery \citep{astroquery_2019}



%\newpage
%\appendix

%\section{Uncertainties of physical properties within the apertures} \label{ap: uncertainties}


\bibliographystyle{aasjournal}
\bibliography{Reference_SFE}



\end{document}